\documentclass[review]{elsarticle}

\usepackage{lineno,hyperref}
\modulolinenumbers[5]
\usepackage{graphicx}
\usepackage{amsmath}
\usepackage{array}
\usepackage{subfig}
\usepackage{fixltx2e}
\usepackage{stfloats}
\usepackage{url}
\usepackage[latin1]{inputenc}
\usepackage[english]{babel}
\usepackage{setspace}
\usepackage{algorithm}
\usepackage{algorithmic}
\usepackage{graphicx}
\usepackage{multirow}
\usepackage{xcolor}
\usepackage{eqnarray}
\usepackage{marvosym}
\usepackage[xcolor]{changebar}
\usepackage{textcomp}
\usepackage{breqn}
\usepackage{tabu}
\def\BibTeX{{\rm B\kern-.05em{\sc i\kern-.025em b}\kern-.08em
		T\kern-.1667em\lower.7ex\hbox{E}\kern-.125emX}}
\usepackage{amssymb}
\usepackage{amsthm}
\usepackage{ wasysym }
\usepackage{extarrows}

\makeatletter
\def\blfootnote{\gdef\@thefnmark{}\@footnotetext}
\makeatother

\begin{document}
\begin{frontmatter}
\title{A Functional Block Decomposition Method for Automatic Op-Amp Design}

	\author{Inga Abel}
\ead{inga.abel@tum.de}

\author{Maximilian Neuner}
\ead{maximilian.neuner@tum.de}

\author{Helmut Graeb}
\ead{helmut.graeb@tum.de}

\address{Technical University of Munich, Arcisstr. 21, 80333 Munich}

\begin{abstract}
This paper presents a method to decompose an op-amp into its functional blocks. The method is able to recognize functional blocks on a high level of abstraction as loads or amplification stages which have a large set of possible structural implementations. The paper presents a hierarchical library of functional blocks. With every hierarchy level, the structural representation of the functional blocks becomes more variable while its function emerges. We use the hierarchical order to automatically  compute the functional decomposition of an op-amp given as a flat netlist.  Experimental results  illustrate the method. 
The functional block decomposition enables a comprehensive formalization of design knowledge for computer-aided design of op-amps.  Applications to circuit sizing and structural synthesis of op-amps are presented. 
\end{abstract}

\end{frontmatter}

\linenumbers

\blfootnote{This document is the results of the research project ``Verification and synthesis of structural properties of analog/mixed-signal circuits by Constraint Programming exemplified by ESD and level shifter circuits'' (GR 1190/6-1) funded by the German Research Foundation (DFG).}

\section{Introduction}

Operational amplifiers are the fundamental building blocks of analog/mixed-signal circuits. They mean to analog design, as some say, what inverters mean to digital design. While inverters and the complete library of cells for digital circuits are designed almost fully automatically, op-amps are still designed mostly manual till now.
This work presents a new approach to the old, but yet unsolved problem of a complete structural and functional representation of op-amps.
As confirmed in \cite{MultiStageOpAmpDesign}, op-amp stage recognition is a hard task.
To tackle this problem, a new hierarchical representation of functional blocks in op-amps is developed in this paper.
 The  method  achieves a complete recognition of all stages in an op-amp, their loads, transconductances and biases.
 The formalized computer-oriented description of the op-amp structure allows a development of new approaches on  major parts on analog design automation (e.g., sizing \cite{ABNG20c}, structural synthesis \cite{ABNG20d}). 
 
 Many types of structural libraries were invented to automate the analog design process. They consist of whole topologies of basic analog circuits \cite{GeometricProgrammingAppFormular, IntegerBasedTopologieSelecting, AGenericTopologySelectionMethodForAnalogCircuits}, predefined modules of, e.g., amplification stages or bias circuits having a certain transistor structure \cite{GeneratorBasedApproachForAnalogCircuitAndLayoutDesignAndOptimization}, 
or basic building blocks being either single devices  with additional self connections as in \cite{AGraphGrammarBasedApproachToAutomatedMultiObjectiveAnalogCircuitDesign, TrustworthyGeneticProgrammingBasedSynthesisOfAnalogCircuitTopologiesUsingHierarchicalDomainSpecificBuildingBlocks} or transistor structures as, e.g., current mirrors, differential pairs   \cite{SizingRuleMethodeForAnalogIntegratedCircuitDesign, Massier2008, FEATS, AnAutomatedTopologySynthesisFrameworkForAnalogIntegratedCircuits}.
The libraries were developed for different purposes. Some are used for topology synthesis \cite{IntegerBasedTopologieSelecting, AGraphGrammarBasedApproachToAutomatedMultiObjectiveAnalogCircuitDesign, TrustworthyGeneticProgrammingBasedSynthesisOfAnalogCircuitTopologiesUsingHierarchicalDomainSpecificBuildingBlocks, FEATS, AnAutomatedTopologySynthesisFrameworkForAnalogIntegratedCircuits,AGenericTopologySelectionMethodForAnalogCircuits}. Other were developed to generate constraints for sizing and layout  \cite{Massier2008}  or to synthesize layouts \cite{GeneratorBasedApproachForAnalogCircuitAndLayoutDesignAndOptimization}.

The libraries come with  disadvantages. As there exist thousands of topology variants for op-amps, libraries containing structural defined topologies or modules as in \cite{GeometricProgrammingAppFormular, IntegerBasedTopologieSelecting,GeneratorBasedApproachForAnalogCircuitAndLayoutDesignAndOptimization, AGenericTopologySelectionMethodForAnalogCircuits} do not support all variants. Adding topologies or modules to the libraries often comes with a high set-up effort. Basic building block libraries, e.g., \cite{AGraphGrammarBasedApproachToAutomatedMultiObjectiveAnalogCircuitDesign, TrustworthyGeneticProgrammingBasedSynthesisOfAnalogCircuitTopologiesUsingHierarchicalDomainSpecificBuildingBlocks, FEATS, AnAutomatedTopologySynthesisFrameworkForAnalogIntegratedCircuits, Massier2008} comprise topology variations. However, used for circuit synthesis, they include  impractical and redundant topologies in the process taking up unnecessary computation time. 
When they are used for sizing and layout generation not all necessary constraints are generated. The methods, e.g. \cite{Massier2008}, are  not able to recognize the load of a first stage, making the usage of other methods necessary \cite{MARS}.
To include more building blocks being important for op-amp design in basic libraries, attempts were made using unsupervised learning algorithms \cite{AnalogCircuitTopologicalFeatureExtractionWithUnsupervisedLearningOfNewSubStructures, Date2021}. However, these methods are still fragile and might not recognize all building blocks correctly.

In this paper, a method is presented which tackles the described disadvantages. It fills in the gap between libraries based on basic building blocks \cite{SizingRuleMethodeForAnalogIntegratedCircuitDesign, Massier2008, FEATS, AnAutomatedTopologySynthesisFrameworkForAnalogIntegratedCircuits} and libraries containing whole topologies \cite{IntegerBasedTopologieSelecting, GeometricProgrammingAppFormular}. We achieve this by capturing all building blocks the full way up to a complete op-amp.
In the frequently cited papers \cite{SizingRuleMethodeForAnalogIntegratedCircuitDesign,Massier2008}, the so-called sizing rules method systematically captured the structures and constraints from differential pairs and current mirrors up to a differential stage. This work additionally captures the load, transconductance and bias of amplification stages as well as the amplification stages themselves and their bias.  A new approach on describing building blocks in op-amps is developed, giving them a well-defined functional and structural description.
While the  structure of a current mirror was relatively easy to establish, capturing the structure of loads, biases and amplification stages is much more complicated. In particular, while the function of these building blocks is well known they are implemented by a large variety of structures. 

With the new method on formalizing existing structural knowledge of op-amps presented in this paper, new approaches to sizing and structural synthesis of op-amps are developed (Sec. \ref{sec:application}, detailed information \cite{ABNG20c,ABNG20d}). Due to the complete recognition of relevant structures in op-amps, the new methods are able to capture op-amp design knowledge presented in a large number of design books, e.g.,  \cite{Allen, LakerSansen, Sansen2007, AnalogIntegratedCircuitDesign, GrayMeyer} to a new extent of completeness using a new systematic presentation making it usable for  computer-aided op-amp design.
The design methods are positioned between classical design methods with more or less fixed design plans   \cite{DEGR87,ELPE89,HARC89,KOSG90a,GeometricProgrammingAppFormular,PDLL01}, which have to be set up for each new op-amp, and simulation-based numerical optimization approaches \cite{ANGW94,OCRC96,LIFG11,MARS,BADP15,PRSF19,CPML19}, which need a CAD tool setup and value seeding for each netlist. It is also positioned between optimization-based structural synthesis approaches \cite{FEATS,VariationAwareStructuralSynthesisOfAnalogCircuitsViaHierarchicalBuildingBlocksAndStructuralHomotopy,  AutomaticSynthesisAndSizingGeneticProgramming,AnAutomatedTopologySynthesisFrameworkForAnalogIntegratedCircuits, AGraphGrammarBasedApproachToAutomatedMultiObjectiveAnalogCircuitDesign} that create a large number of variants some of them being impractical or redundant, and approaches without involvement of optimization that investigate only a very small number of variants \cite{NovelCircuitTopologySynthesisMethodUsingCircuitFeatureMiningAndSymbolicComparison,IntegerBasedTopologieSelecting}.

The remainder of the paper is organized as follows: Sec.~\ref{sec:functionalBlocks} gives a general overview of functional blocks in op-amps. Detailed explanations of the individual functional block are given in Secs. \ref{sec:DeviceLevelFunctionalBlock} - \ref{sec:FunctionalBlockLevel4}. The algorithms to recognize the blocks in an op-amp are presented in Sec. \ref{sec:functionalBlockAnalysis}. In Sec. \ref{sec:experimentalResults}, we discuss the corresponding experimental results. Sec. \ref{sec:application} shows the application of the functional block decomposition method on circuit synthesis and sizing. A conclusion and an outlook on future work are presented in Secs. \ref{sec:conclusion}.

\section{Functional Blocks in Op-Amps}\label{sec:functionalBlocks}

 \begin{figure}\centering
	\includegraphics[width=0.99\linewidth]{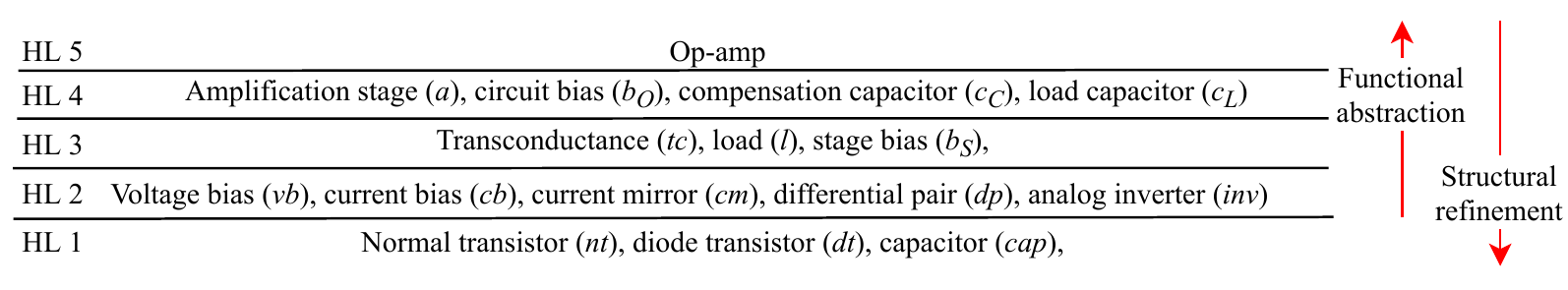}
	\caption{Hierarchical library of functional blocks in op-amps}
	\label{fig:functionalBlockPyramid}
\end{figure}

An op-amp can be hierarchically decomposed into functional blocks. Fig.~\ref{fig:functionalBlockPyramid} shows this decomposition. Starting with one functional block at the highest hierarchy level (HL 5), the number of functional blocks per level increases till on the lowest level (HL 1) every device of the circuit forms a functional block of its own. On HL 1, every functional block is represented by one uniquely definable device composition. However, this level does not give any information about the functional task  a device fulfills in the op-amp, e.g. amplification, stabilization, biasing.
This functional assignment is given at HL 3 and HL 4. At these levels, the function of every functional block is uniquely definable. However, the structural description is not unique as e.g. different types of amplification stages exist.
On HL 2, neither the structural nor the functional definition of the functional blocks is unique. However, their structural complexity is less then those of the functional blocks of HL 3 and HL 4.

\begin{figure} [t] \centering
	\subfloat[Functional blocks on HL 1 - HL 2]{
		\label{fig:symmetricalOPAmpHL12}
		\includegraphics[width=0.45\linewidth]{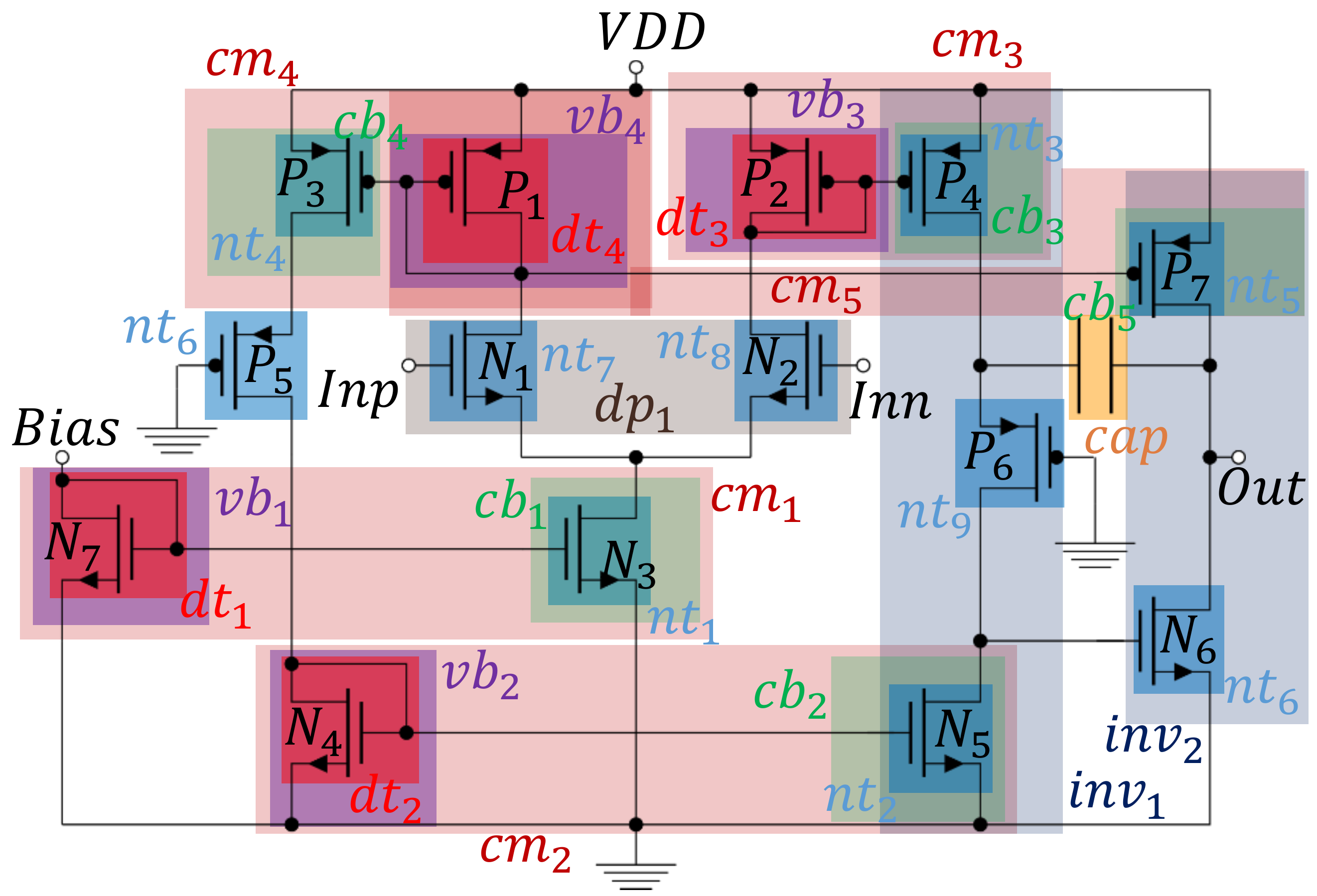}
	}%
	\qquad
	\subfloat[Functional blocks on HL 3 - HL 4]{
		\label{fig:symmetricalOPAmpHL34}
		\includegraphics[width=0.45\linewidth]{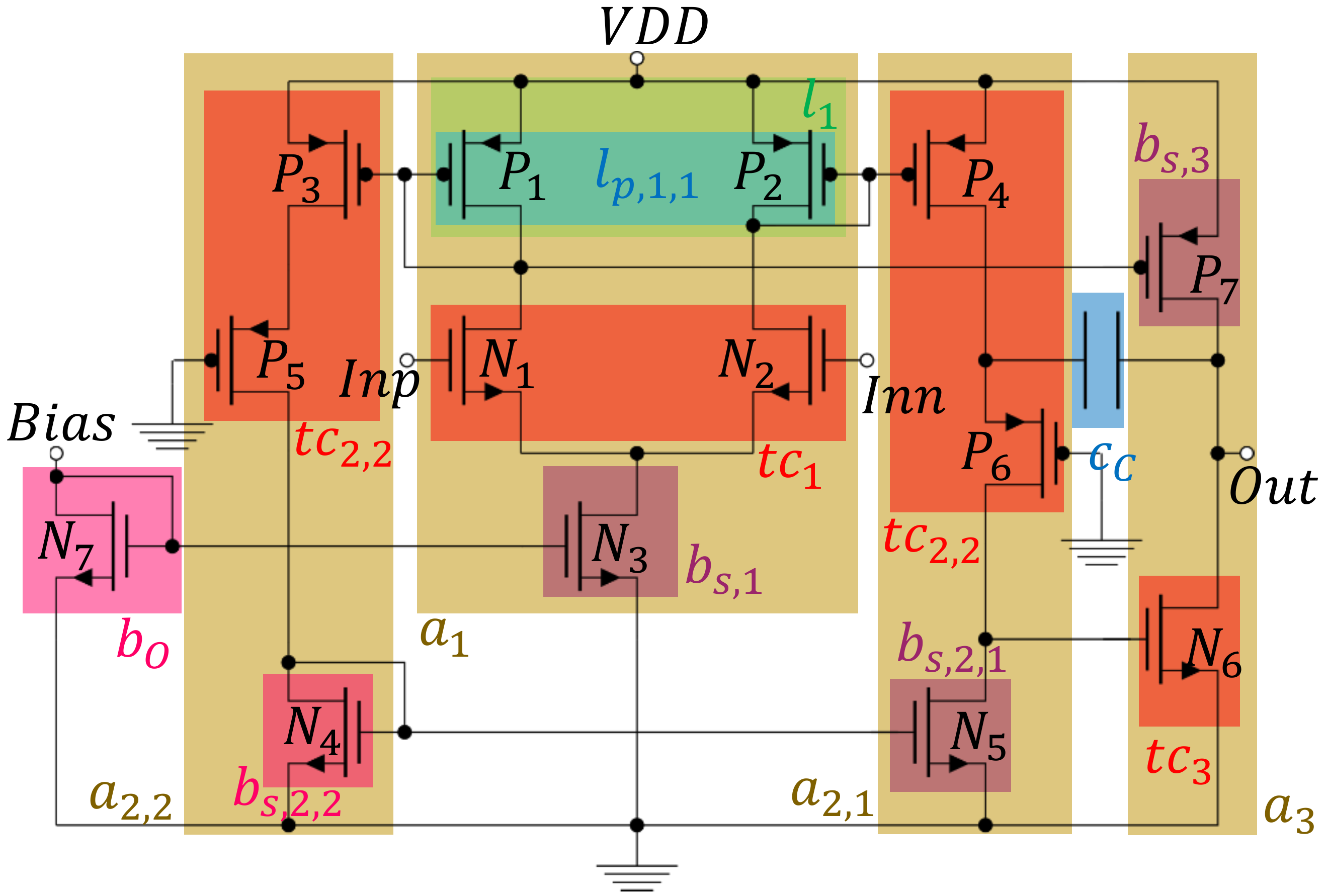}
	} \qquad
	\caption{Symmetrical op-Amp with high PSRR \protect \cite{LakerSansen} (Definitions of abbreviations in Fig. \protect \ref{fig:functionalBlockPyramid})}
	\label{fig:symmetricalOpAmpWithSecondStage}
\end{figure}

Fig. 2 shows as an example the functional blocks in a symmetrical op-amp. For HL 1 and HL 2 (Fig \ref{fig:symmetricalOPAmpHL12}), all functional blocks of the same type have similar device compositions. This is different for  HL 3 and HL 4 (Fig \ref{fig:symmetricalOPAmpHL34}). The composition of the first stage $a_1$ differs highly from the composition of the other stages.
On HL 1, HL 3, and HL 4, all devices can only be part of one functional block on a level. $N_5$ is a normal transistor $nt_2$ on level 1, a stage bias $b_{2,1}$ on level 3 and part of an amplification stage $a_{2,1}$ on level 4.
On HL 2, a device can be part of more than one functional block. On this level, $N_5$ is  a current bias $cb_2$, part of a current mirror $cm_2$ and part of an analog inverter $inv_1$.

We define the set of  functional blocks $\mathcal{X}$ in an op-amp as:
\begin{equation}
\begin{split}
\mathcal{X} = \{x_{k} | & k = 1, 2, .., |\mathcal{X}|\}
\end{split}
\end{equation}

A functional block $x_k$ in a circuit consists either of a basic  device $d_k$ of type $d_k.type  \in \{t,c\}$, where $t$ is referring to transistors and $c$ to capacitors, or of other functional blocks $x_{k,1}, .. , x_{k,n}$ of individual types \linebreak $x_{k,l}.type \in \{dt, nt, cap, vb, cb, cm, dp, inv, g, l ,b, a, c_L, c_C\}$ (see Fig. \ref{fig:functionalBlockPyramid}):
\begin{equation}
\forall_{x_k\in \mathcal{X}} ~ (x_k = \{d_k\} \vee x_k = \{x_{k,1}, .. ,x_{k,n} \})
\end{equation}
The subset $\mathcal{X}_{j} \subseteq \mathcal{X}$ contains all functional blocks  $x_k \in \mathcal{X}$ with $x_k.type = j$. 

To describe the connections between functional blocks, every functional block $x_{k}$ is assigned a set of pins $P_{x_{k}}$, which are connected to the nets of the circuit:
\begin{equation}
P_{x_{k}} = \{x_{k}.p_{l}| l = 1, 2, .. |P_{x_{k}}| \}
\end{equation}
As all functional blocks consist at the end of devices,  the pins $P_{x_k}$ of a functional block $x_k$ refer to certain device pins.

A connection of two functional blocks $x_k, x_l$ over any net with the pins $x_k.p_y, x_l.p_z$ is described by:
\begin{equation}
x_k.p_y \leftrightarrow x_l.p_z
\end{equation}
To describe that two pins  $x_k.p_y, x_l.p_z$ are not allowed to be connected by any net, the following notation is used:
\begin{equation}
	x_k.p_y \nleftrightarrow x_l.p_z
\end{equation}

$x_k.\Phi$ denotes the substrate type of a functional block $x_k$, with following naming convention:
\begin{equation}
x_k.\Phi \in \{ \Phi_n, \Phi_p, \Phi_u\}, ~~\text{for n-, p-, or mixed-doping}
\end{equation}
A functional block $x_k$ has mixed-doping if it consists of transistors with different doping. 

In the following, structural and functional definitions will be given for all functional block types in Fig. \ref{fig:functionalBlockPyramid}.
Please note that all examples shown for NMOS transistors, hold analogously for PMOS transistors.

\section{Functional Blocks on Hierarchy Level 1}\label{sec:DeviceLevelFunctionalBlock}

 \begin{figure}[t]\centering
	\includegraphics[width=0.6\linewidth]{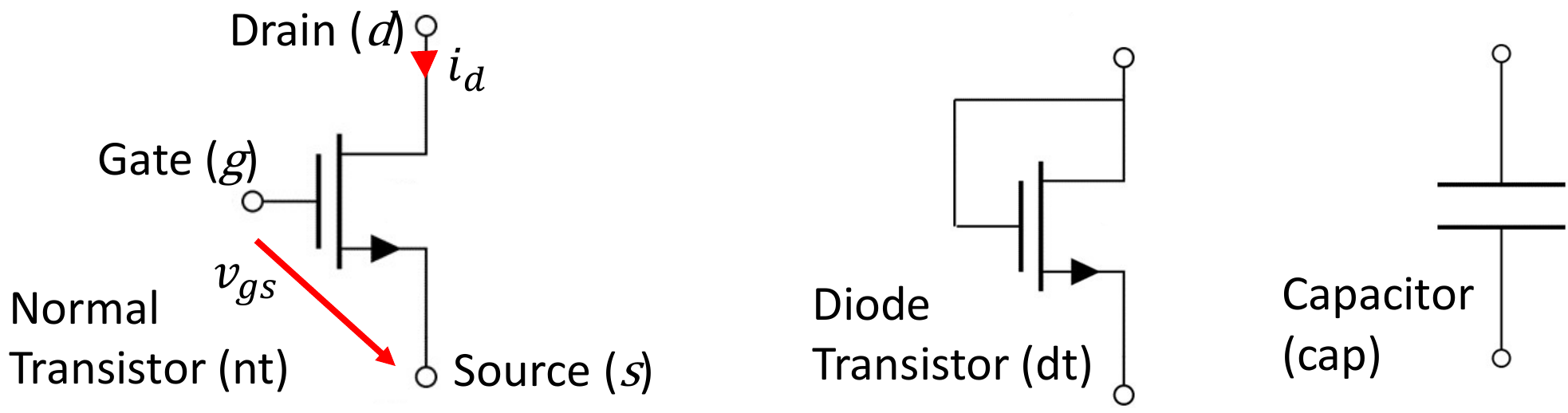}
	\caption{Functional blocks on level 1}
	\label{fig:deviceLevelFunctionalBlocks}
\end{figure}

Fig. \ref{fig:deviceLevelFunctionalBlocks} shows three different device level functional blocks. In addition to capacitors, we define and use two different functional block types for  transistors:

\subsection{Normal Transistor}

{\em Function:}
A normal transistor $nt_k$ establishes a relation between the the voltage potential $v_g$ at the gate of the transistor and the  current at the drain $i_d$ of the transistor. Either $v_g$ or $i_d$ is the control factor.

{\em Structure:}
A normal transistor $nt_k$ is a transistor $d_k$ without  any self connections.
\begin{equation}\label{eq:nt}
\begin{split}
x_k =& \{d_k\} \wedge d_k.type = t  \wedge d_k.d  \nleftrightarrow d_k.s \wedge d_k.d  \nleftrightarrow d_k.g \\ &\wedge d_k.g  \nleftrightarrow d_k.s  
 \Leftrightarrow x_k.type = nt
\end{split}
\end{equation}

\subsection{Diode Transistor} 
{\em Function:}
A diode transistor $dt$ converts its drain-source current $i_d$  into  a stable gate-source voltage $v_{gs}$.

{\em Structure:}
A diode transistor $dt_k$ is a transistor $d_k$, whose drain $d_k.d$ is connected to its gate $d_k.g$:
\begin{equation}\label{eq:dt}
\begin{split}
x_k = \{d_k\}  \wedge d_k.type =  t \wedge d_k.d  \leftrightarrow d_k.g \wedge d_k.d  \nleftrightarrow d_k.s  
\Leftrightarrow x_k.type = dt
\end{split}
\end{equation}

\section{Functional Blocks on Hierarchy Level 2}\label{sec:FunctionalBlockLevel2}

The majority of the functional blocks on  HL 2 consist of transistor stacks. Typical transistor stacks are shown in Fig.~\ref{fig:voltageBiases},~\ref{fig:currentBiases}. A transistor stack $ts_k$ is a set of 1-3 transistors having the same doping and a drain-source connection i.e., the drain of a lower transistor in the stack $x_{k,m}.d$ is connected with the source of the next higher transistor $x_{k,m+1}.s$. Higher transistor gates are not allowed to be connected to drains of lower transistors. Drains of higher transistors are not allowed to be connected to lower transistor sources. 
\begin{equation}\label{eq:ts}
\begin{split}
 x_k =&\{x_{k,1}, .. , x_{k,n} |n = |x_k| \wedge n \leq 3\} \wedge x_{k} \subset (\mathcal{X}_{nt} \cup \mathcal{X}_{dt})\wedge x_{k,m}.d \leftrightarrow x_{k,m+1}.s \\
&\wedge x_{k,m}.\Phi = x_{k,m+1}.\Phi \wedge x_{k,m+1}.g
\nleftrightarrow x_{k,m}.d\wedge x_{k,m+1}.d \nleftrightarrow x_{k,m}.s\\ &\Leftrightarrow x_k.type = ts
\end{split}
\end{equation} 
The usual number of transistors in a stack is 1-2. 
For the pins in a transistor stack, the following naming convention will be used. The source not connected to any drain in the stack $ts_{k}.s_1$ is the source $ts_k.s$ of the transistor stack. The drain not connected to any source of the stack $ts_{k}.d_n$ is the drain $ts_k.d$ of the transistor stack. If all drains or sources of the stack must be considered, numbering will be used.
The definition of the transistor stack is in the following used to describe the functional blocks in detail.

\subsection{Voltage Bias}\label{sec:VoltageBias}
 \begin{figure}\centering
	\includegraphics[width=0.8\linewidth]{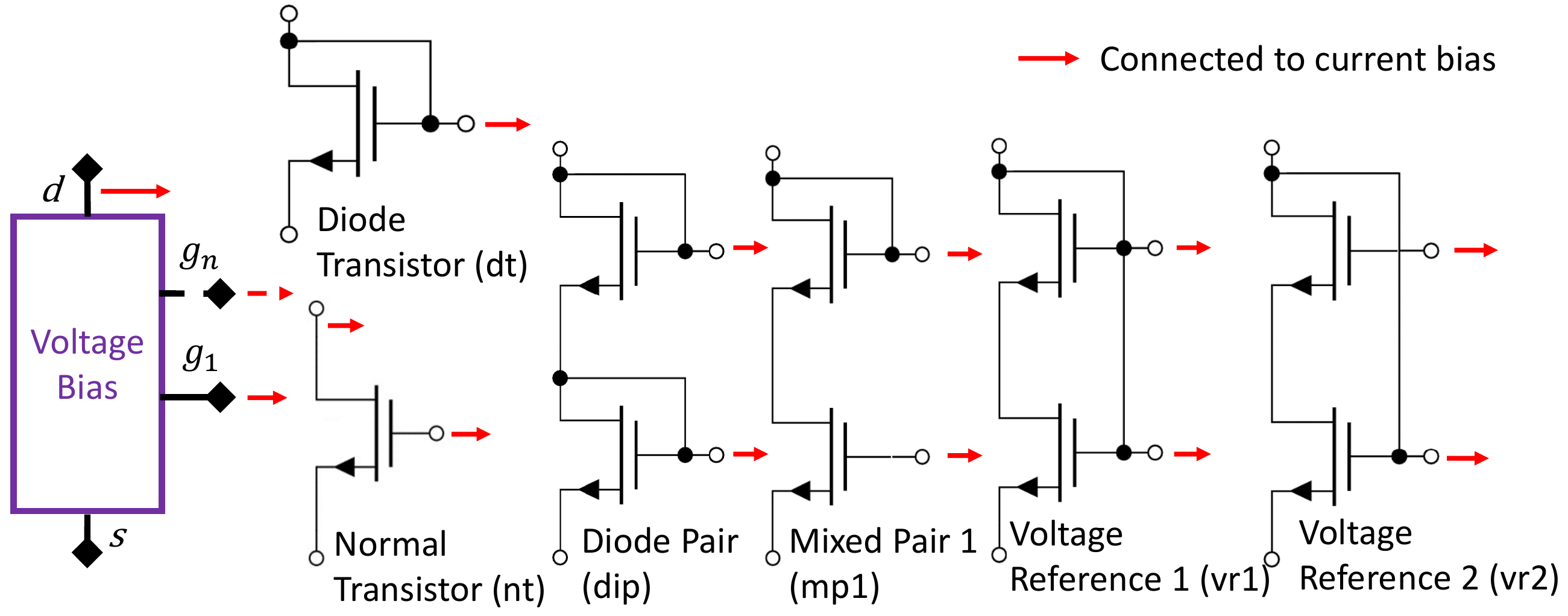}
	\caption{Voltage bias and  variants with stacks of 1 or 2 transistors (dashed lines: optional pins and functional blocks)}
	\label{fig:voltageBiases}
\end{figure} 
{\em Function:}
A voltage bias $vb_k$  converts its drain current $i_d$ into a stable gate-source voltage $v_{gs}$  applied to a gate of a current bias.

{\em Structure:}
A voltage bias $vb_k$ (Fig. \ref{fig:voltageBiases}) consists of a transistor stack \linebreak $x_k, x_k.type =ts,$ with the gates of its  devices $x_{k}.g_l$ connected to gates of a current bias $cb_v$ with same doping as $vb_k$. Additionally, the drain of the stack $x_{k}.d = x_{k,n}.d$ must be connected to a gate $cb_v.g_m$ of $cb_v$. For every gate $x_{k}.g_j$  in the stack, exactly one gate-drain connection with another transistor $x_y \in (\mathcal{X}_{nt} \cup \mathcal{X}_{dt})$ of same doping exists. This transistor $x_y$ must be part of a voltage or current bias $x_z \in (\mathcal{X}_{vb} \cup \mathcal{X}_{cb})$ but not necessarily of $vb_k$ or $cb_v$ itself.
\begin{equation}\label{eq:vb}
\begin{split}
&x_k = \{x_{k,1}, .. , x_{k,n} |n = |x_k|\} \wedge x_k.type = ts 
 \wedge \exists_{cb_v} \big[  cb_v.\Phi = x_k.\Phi  \wedge x_{k,n}.d \\
 &\leftrightarrow cb_v.g_m \wedge  \forall_{x_{k}.g_l} 
 [x_{k}.g_l \leftrightarrow cb_v.g_i  ]\big]   \wedge \forall_{x_k.g_j} \big[  \exists!_{x_y \in (\mathcal{X}_{nt} \cup \mathcal{X}_{dt})} [(x_y.\Phi = x_k.\Phi  \\
 &\wedge  x_y.d \leftrightarrow x_k.g_j)  \longleftrightarrow \exists!_{x_z \in (\mathcal{X}_{vb} \cup \mathcal{X}_{cb})} x_y \in x_z ] \big]
 \Leftrightarrow x_k.type = vb
\end{split}
\end{equation}

\subsection{Current Bias}\label{sec:currentBias}

 \begin{figure}\centering
	\includegraphics[width=0.65\linewidth]{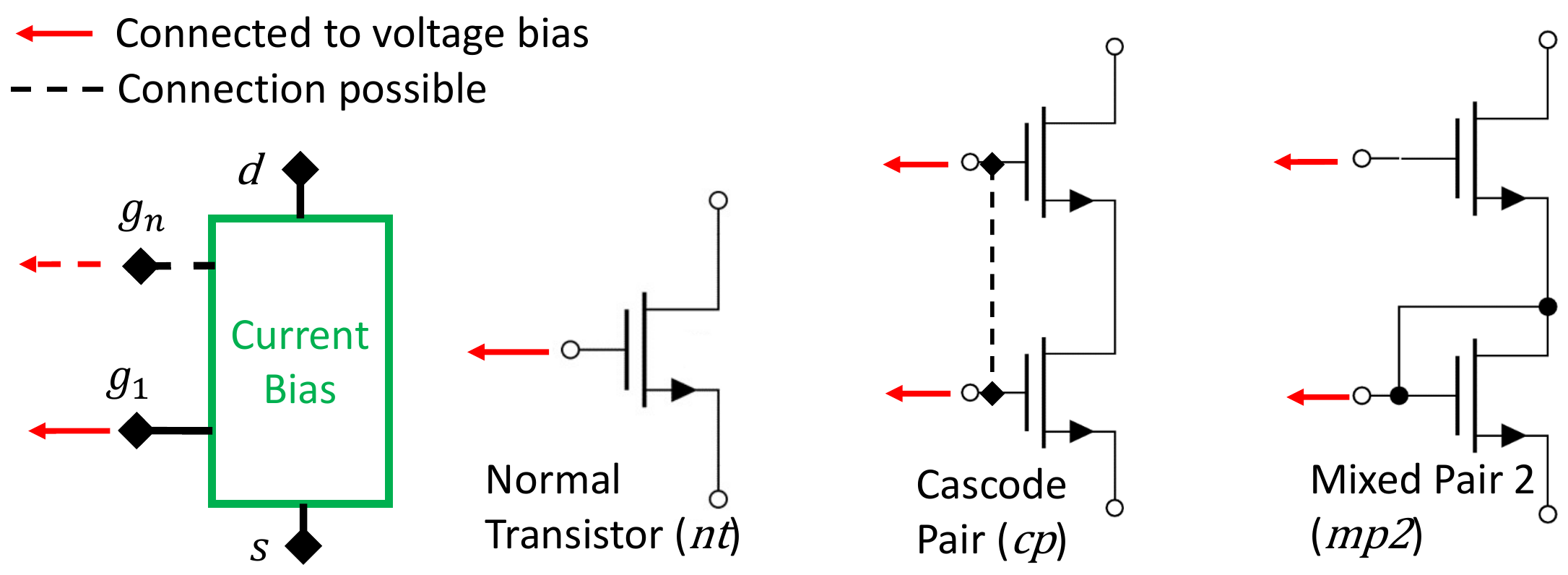}
	\caption{Current bias and variants with stacks of 1 or 2 transistors (dashed lines: optional) }
	\label{fig:currentBiases}
\end{figure} 

{\em Function:}
A current bias $cb_k$ converts the voltage potential $v_g$ applied at its gates into a drain current $i_d$.

{\em Structure:}
A current bias $cb_k$ (Fig. \ref{fig:currentBiases}) consists typically of a stack of normal transistors, which might have a gate connection. In rare cases, the normal transistor at the bottom is exchanged by a diode transistor. The gates of a current bias $cb_k.g_l$ are  connected to a gate or the upper drain of a voltage bias ($vb_v.g_l \vee vb_v.d$). The voltage bias must have the same doping. The drain of the upper transistor of a current bias $cb_{k,n}.d = cb_{k}.d$ must not be connected to any gate of a voltage bias $vb_y.g_z$ with same doping.
\begin{equation}\label{eq:cb}
\begin{split}
&x_k  =\{x_{k,1}, .. , x_{k,n} |n = |x_k|\}\wedge x_k.type = ts \wedge   \exists_{vb_v} \big[ vb_v.\Phi = x_k.\Phi \wedge \forall_{x_k.g_l} [x_{k}.g_l \\
&   \leftrightarrow(vb_v.g_l \vee vb_v.d) \big] \wedge  \not\exists_{vb_y} [ vb_y.\Phi = x_k.\Phi \wedge x_{k,n}.d \leftrightarrow vb_y.g_z] \big]   \Leftrightarrow x_k.type = cb
\end{split}
\end{equation} 

\subsection{Current Mirror}\label{sec:currentMirror}

 \begin{figure}
	\includegraphics[width=0.9\linewidth]{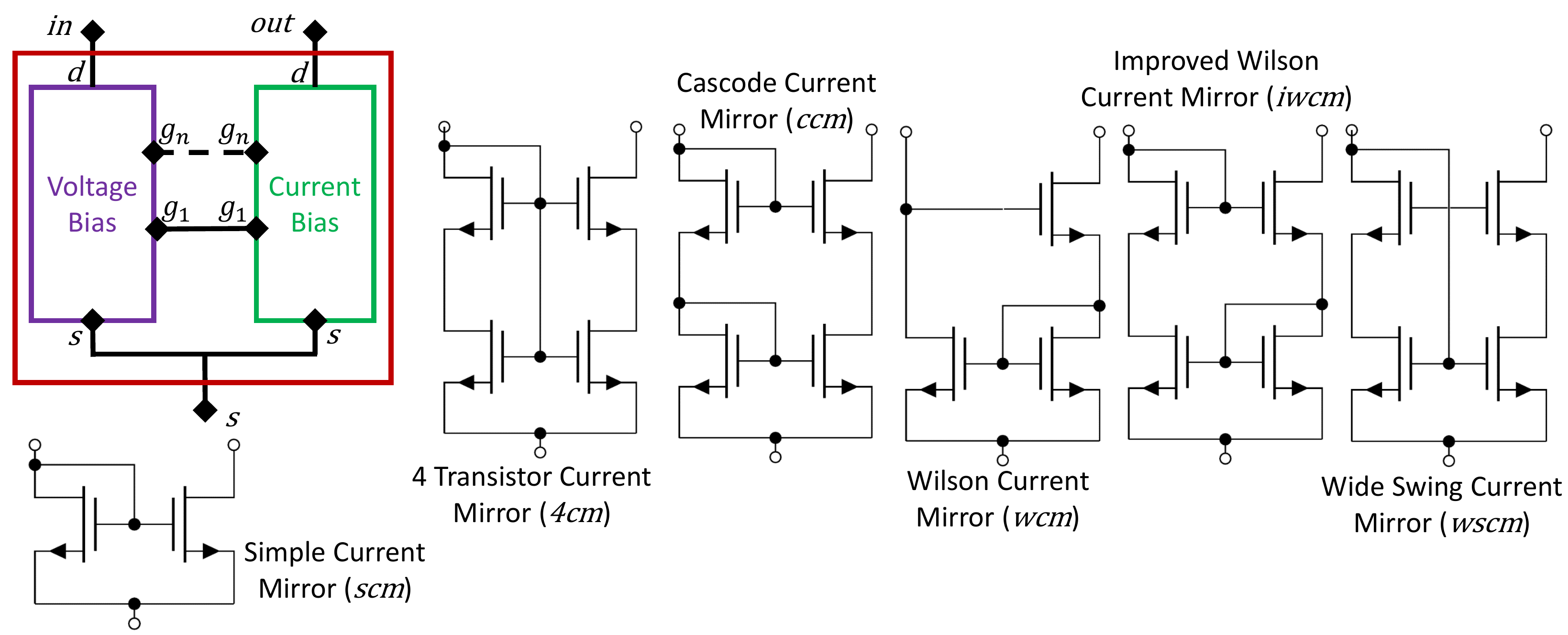}
	\caption{Current mirror and examples (dashed lines: optional) }
	\label{fig:currentMirror}
\end{figure} 
{\em Function: }
A current mirror $cm_k$ provides a current by the current bias, specified by the devices sizes and the current of the voltage bias.

{\em Structure:}
Fig. \ref{fig:currentMirror} shows the structural definition of a current mirror and examples.
To form a current mirror $cm_k$, voltage bias $vb_k$ and current bias $cb_k$ must have equal doping and a source connection.  The gates of the voltage bias $vb_k.g_l$ must be connected to the gates $cb_k.g_l$ of the current bias. The uppermost drain of the voltage bias $vb_k.d$ must be connected to a gate of the current bias $cb_k.g_m$. All gates of the voltage bias $vb_k.g_j$ with exception of the upper most one $vb_k.g_{|vb_k|}$ must have a connection to a drain of either voltage or current bias $x_y.d_z|_{x_y \in \{vb_k, cb_k\}}.$
\begin{equation}\label{eq:cm}
\begin{split}
&x_k = \{vb_{k}, cb_{k}\}  \wedge vb_k.\Phi = cb_k.\Phi \wedge vb_k.s \leftrightarrow cb_k.s \wedge  (vb_k.g_l \leftrightarrow cb_k.g_l)|_ {l \leq |vb_k|}\\
&  \wedge \exists_{cb_k.g_m} [cb_k.g_m \leftrightarrow vb_k.d]   \wedge \forall_{vb_k.g_j \in P_{vb_k} \setminus \{vb_k.g_{|vb_k|}\} }\big[ \exists!_{x_y.d_z \in P_{x_y}|_{x_y \in \{vb_k, cb_k\}}} \\ 
&[ vb_k.g_j \leftrightarrow x_y.d_z]    \big]\Leftrightarrow x_k.type = cm
\end{split}
\end{equation}

\subsection{Differential Pair}\label{sec:differentialPair}

{\em Function: }
A differential pair $dp_k$  converts the voltage potentials at the gates of its two input transistor into  amplified drain currents. Equal voltages lead to  equal drain currents.
Depending on the structure, higher amplification gains can be obtained. Cascode versions with four transistors have a higher amplification gain than a simple version with two transistors.

\begin{figure}\centering
	\includegraphics[width=0.99\linewidth]{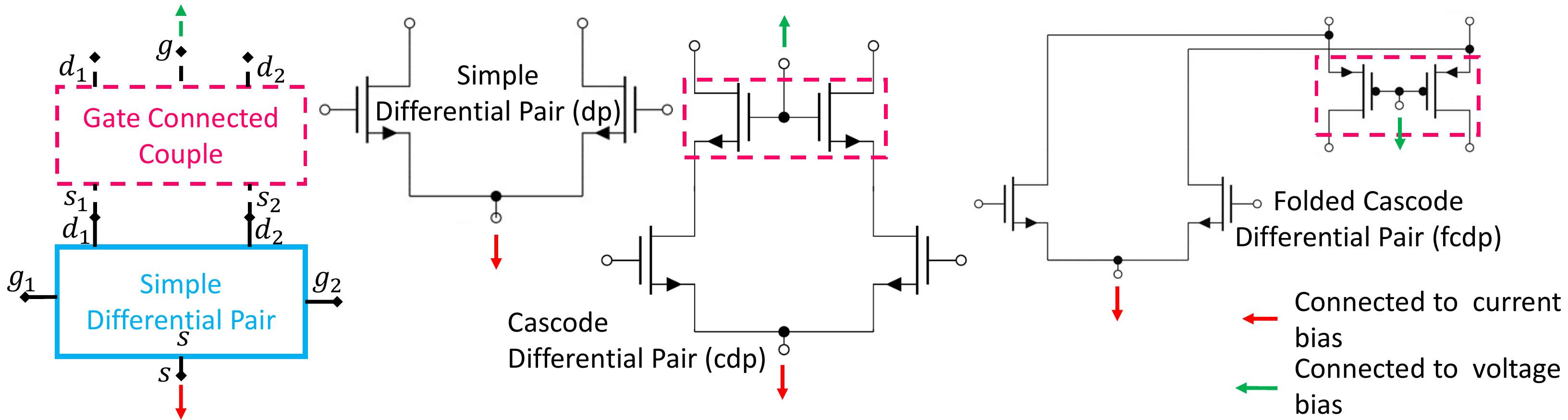}
	\caption{Differential pair and examples
		(dashed lines: optional) 
	}
	\label{fig:differentialPairAndItsVariants}
\end{figure}

{\em Structure:}
Fig. \ref{fig:differentialPairAndItsVariants} shows the structure of a differential pair. The basic structure is the simple differential pair, which can also stand alone. 

A { \em simple differential pair} $dp_k$ consist of two normal transistors $nt_{k,1}, nt_{k,2}$ connected only at their sources. This common source $dp_k.s$ must be connected to a current bias drain $cb_l.d$. The two normal transistors and the current bias must have equal doping.
\begin{equation}\label{eq:dp}
\begin{split}
&x_k = \{nt_{k,1}, nt_{k,2}\}   \wedge  nt_{k,1}.\Phi = nt_{k,2}.\Phi \wedge nt_{k,1}.s \leftrightarrow nt_{k,2}.s \wedge nt_{k,1}.d/g \\
&  \nleftrightarrow nt_{k,2}.d/g\wedge  \exists_{cb_l} [nt_{k,1}.s \leftrightarrow cb_l.d \wedge nt_{k,1}.\Phi = cb_l.\Phi ] \Leftrightarrow x_k.type = dp
\end{split}
\end{equation}

For the cascode version of the differential pair $vdp_k$, a simple differential pair $dp_k$ is connected with its drains to the sources of a {\em gate connected couple} $gcc_k$. These are two normal transistor with same doping  connected at their gates:
\begin{equation}\label{eq:gcc}
\begin{split}
&x_k = \{nt_{k,1}, nt_{k,2} \}  \wedge  nt_{k,1}.\Phi = nt_{k,2}.\Phi\wedge nt_{k,1}.g \leftrightarrow nt_{k,2}.g \\
& \wedge nt_{k,1}.d/s \nleftrightarrow nt_{k,2}.d/s  \Leftrightarrow x_k.type = gcc
\end{split}
\end{equation}
The structural definition of the {\em cascode version of the differential pair $vdp_k$} is:
\begin{equation}\label{eq:vdp}
\begin{split}
x_k = &\{dp_{k}, gcc_{k}\}  \wedge (dp_{k}.d_l \leftrightarrow gcc_{k}.s_l)|_{l=1,2}\wedge dp_{k}.s/g_{1,2} \nleftrightarrow gcc_{k}.g/d_{1,2} \\ 
 &\Leftrightarrow  x_k.type = vdp
\end{split}
\end{equation}

Two types of cascode variants exist having different doping characteristics. In the {\em folded cascode differential pair } $fcdp_k$, $dp_k$ and $ggc_k$ have opposite doping:
\begin{equation}\label{eq:fcdp}
\begin{split}
vdp_k = &\{dp_{k}, gcc_{k}\}  \wedge dp_{k}.\Phi \neq gcc_{k}.\Phi \Leftrightarrow  vdp_k.type = fcdp
\end{split}
\end{equation}
In the {\em cascode differential pair} $cdp_k$, they have equal doping:
\begin{equation}\label{eq:cdp}
\begin{split}
vdp_k = &\{dp_{k}, gcc_{k}\}  \wedge dp_{k}.\Phi = gcc_{k}.\Phi \Leftrightarrow  vdp_k.type = cdp
\end{split}
\end{equation}

\subsection{Analog Inverter}\label{sect:analogInverter}

 \begin{figure}\centering
	\includegraphics[width =0.65\linewidth]{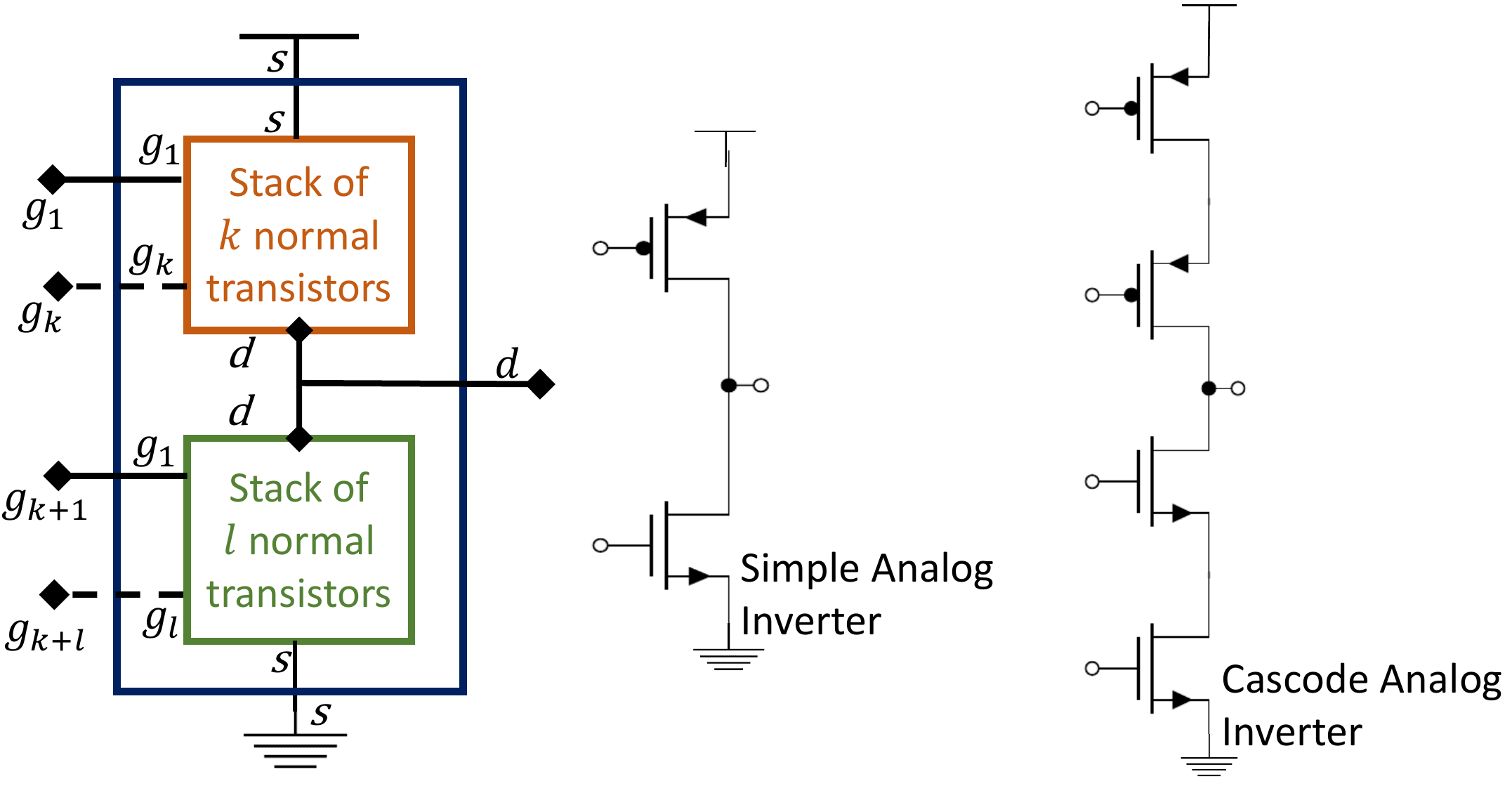}
	\caption{Analog inverter and examples
	(dashed lines: optional)
	}
	\label{fig:inverter}
\end{figure} 
{\em Function: }
An analog inverter inverts and amplifies an input voltage applied at one of its gates. 

{\em Structure:}
An analog inverter $inv_k$ (Fig. \ref{fig:inverter})  consists of two normal transistor stacks $ts_{k,1}, ts_{k,2}$ differing in doping. 
The stacks are connected at their drains $ts_{k,1}.d, ts_{k,2}.d$. The two sources $ts_{k,1}.s, ts_{k,2}.s$ are connected to the supply voltage rail that corresponds to the doping type. Gate-gate, gate-drain and source-source connections between transistors are not allowed.
\begin{equation}\label{eq:inv}
\begin{split}
x_k & = \{ts_{k,1}, ts_{k,2}\}  \wedge (ts_{k,1} \cup ts_{k,2}) \subset \mathcal{X}_{nt} \wedge ts_{k,1}.d \leftrightarrow ts_{k,2}.d 
\wedge ts_{k,1}.\Phi = \Phi_ p   \\
&\wedge  ts_{k,2}.\Phi = \Phi_n\wedge ts_{k,1}.s \leftrightarrow VDD  \wedge  ts_{k,2}.s \leftrightarrow GND 
\\&\wedge \forall_{nt_i,nt_j \in (ts_{k,1} \cup ts_{k,2})} [nt_{i}.g \nleftrightarrow nt_{j}.g \wedge nt_{i}.d \nleftrightarrow nt_{j}.g \wedge nt_{i}.s \nleftrightarrow nt_{j}.s ]\\ 
& \Leftrightarrow x_k.type = inv
\end{split}
\end{equation}

\subsection{Multiple Assignments of Transistors to Functional Blocks} \label{sec:feasibleAndUnfeasibleAssignments}

Hierarchy level 2 is the only level which allows transistors to be part of more than one functional block.
An example  are current mirrors. According to \eqref{eq:cm}, every transistor in a current mirror $t_k \in cm_k$ is at the same time part of a voltage  or current bias $t_k \in (vb_k \cup cb_k)$.
We distinguish between relevant, irrelevant and false multiple assignments. 

\begin{figure} [tb] \centering
	\subfloat[Current biases with gate connected couple]{
		\label{fig:currentBiasesWithGCC}
		\includegraphics[scale = 0.25]{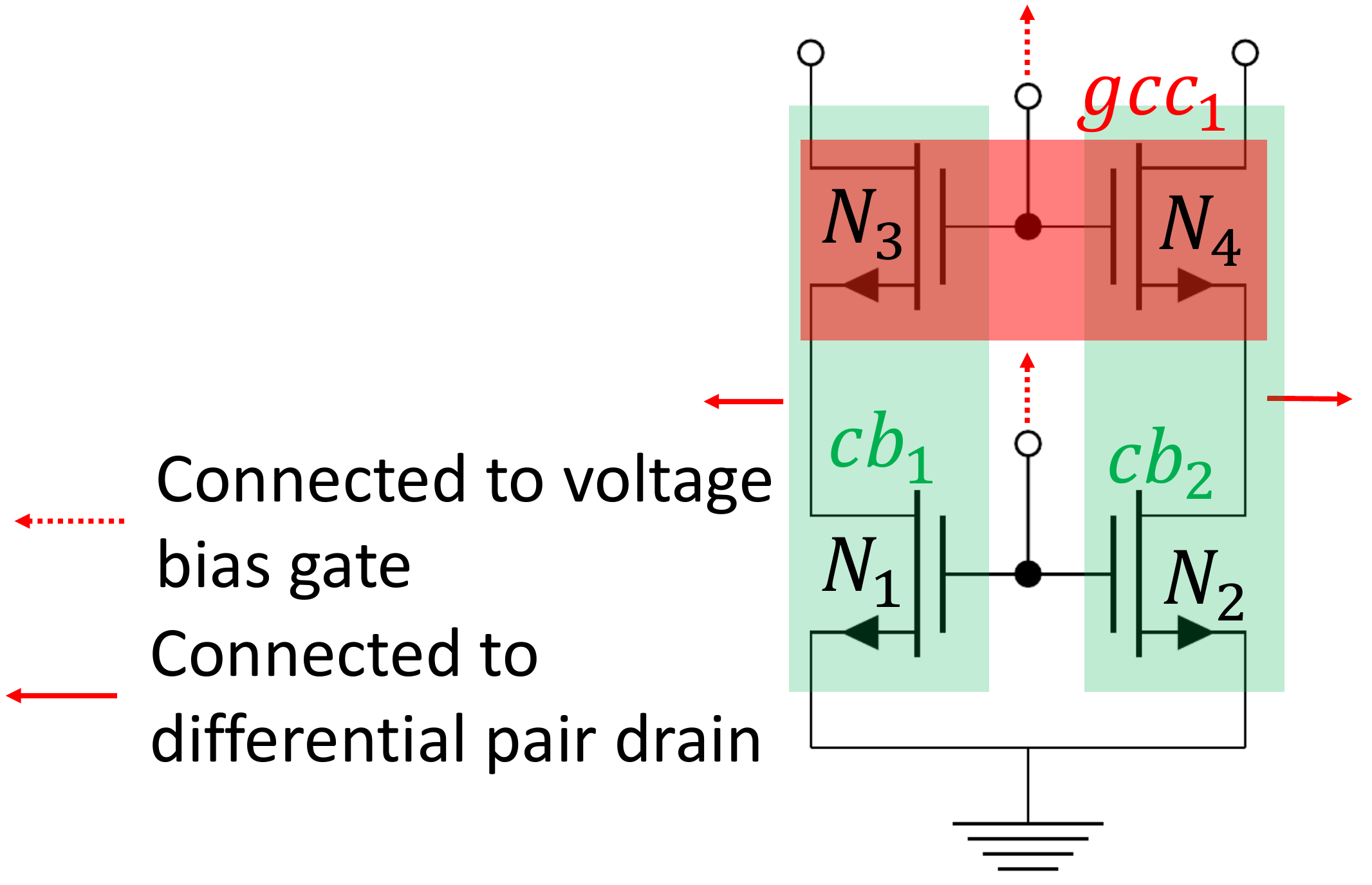}
	}%
	\qquad
	\subfloat[Inverter with current mirror]{
		\label{fig:inverterWithCurrentMirror}
		\includegraphics[scale = 0.25]{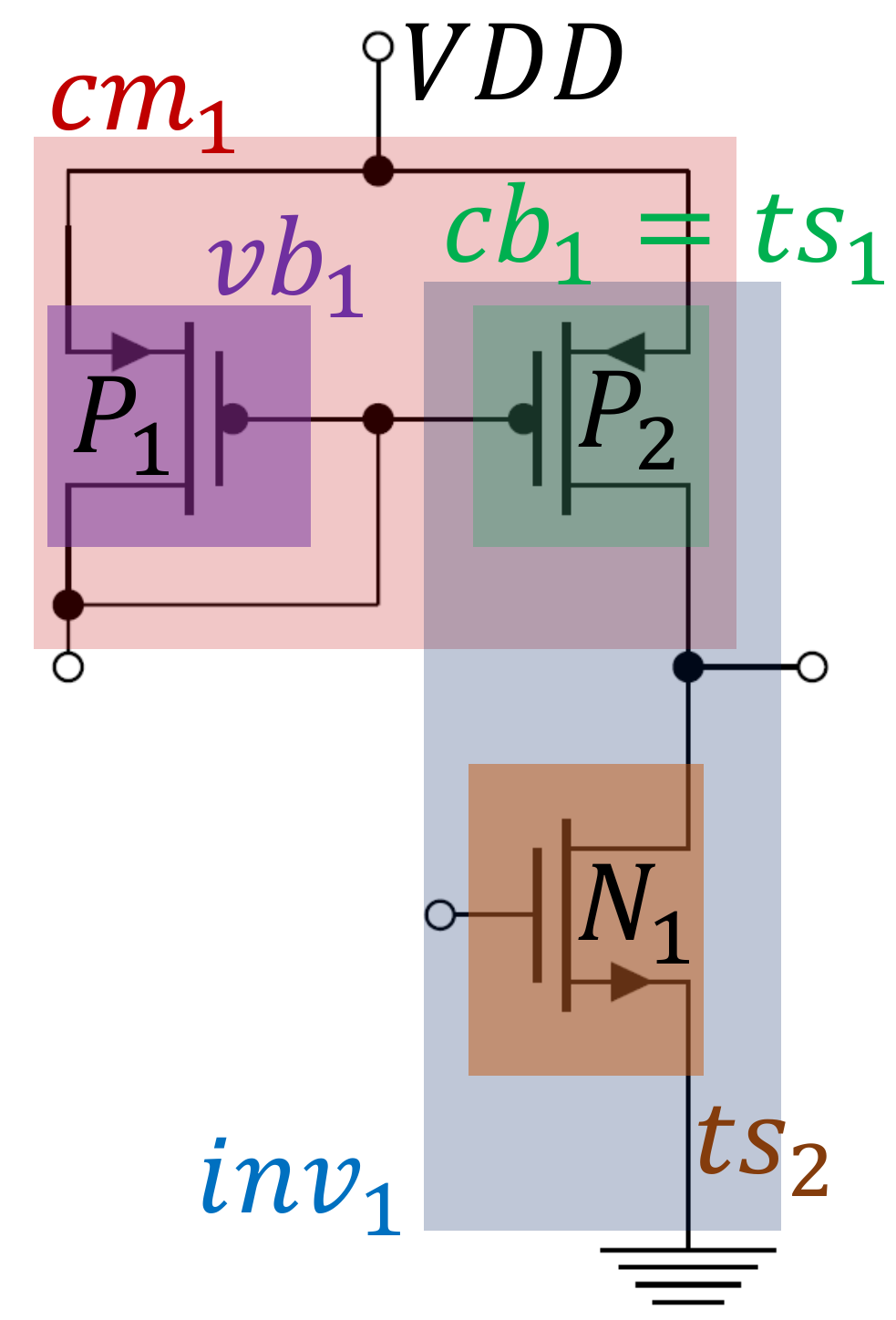}
	}%
	\qquad
	\caption{Relevant multiple assignments}
	\label{fig:feasibleDoubleAssignments}
\end{figure}

{\em Relevant multiple assignments} are circuitry-wise correct and obtain additional  information needed to find other functional blocks. These are the above mentioned double assignments in current mirrors, different current mirrors with the same voltage bias but different current biases forming a current mirror bench, and the two cases shown in Fig.  \ref{fig:feasibleDoubleAssignments}.
Fig. \ref{fig:currentBiasesWithGCC} shows a gate connected couple \eqref{eq:gcc} $gcc_1$ in two current biases $cb_1,cb_2$. The gate connected couple must be part of a cascode version of a differential pair $vdp_1$ \eqref{eq:vdp}. The transistors $N_3,N_4$ are therefore part of $gcc_1,vdp_1$ and $cb_1$ or $cb_2$.
Fig. \ref{fig:inverterWithCurrentMirror} shows an analog inverter with a transistor stack $ts_1 = \{P_2\}$ that is also part of a current mirror $cm_1$. Thus, $P_2$ is part of $inv_1, cm_1$ and  the current bias $cb_1$ in $cm_1$.

\begin{figure} [tb] \centering
	\subfloat[Simple current mirror in a cascode version]{
		\label{fig:cascodeCurrentMirrorWithSCM}
		\includegraphics[scale = 0.3]{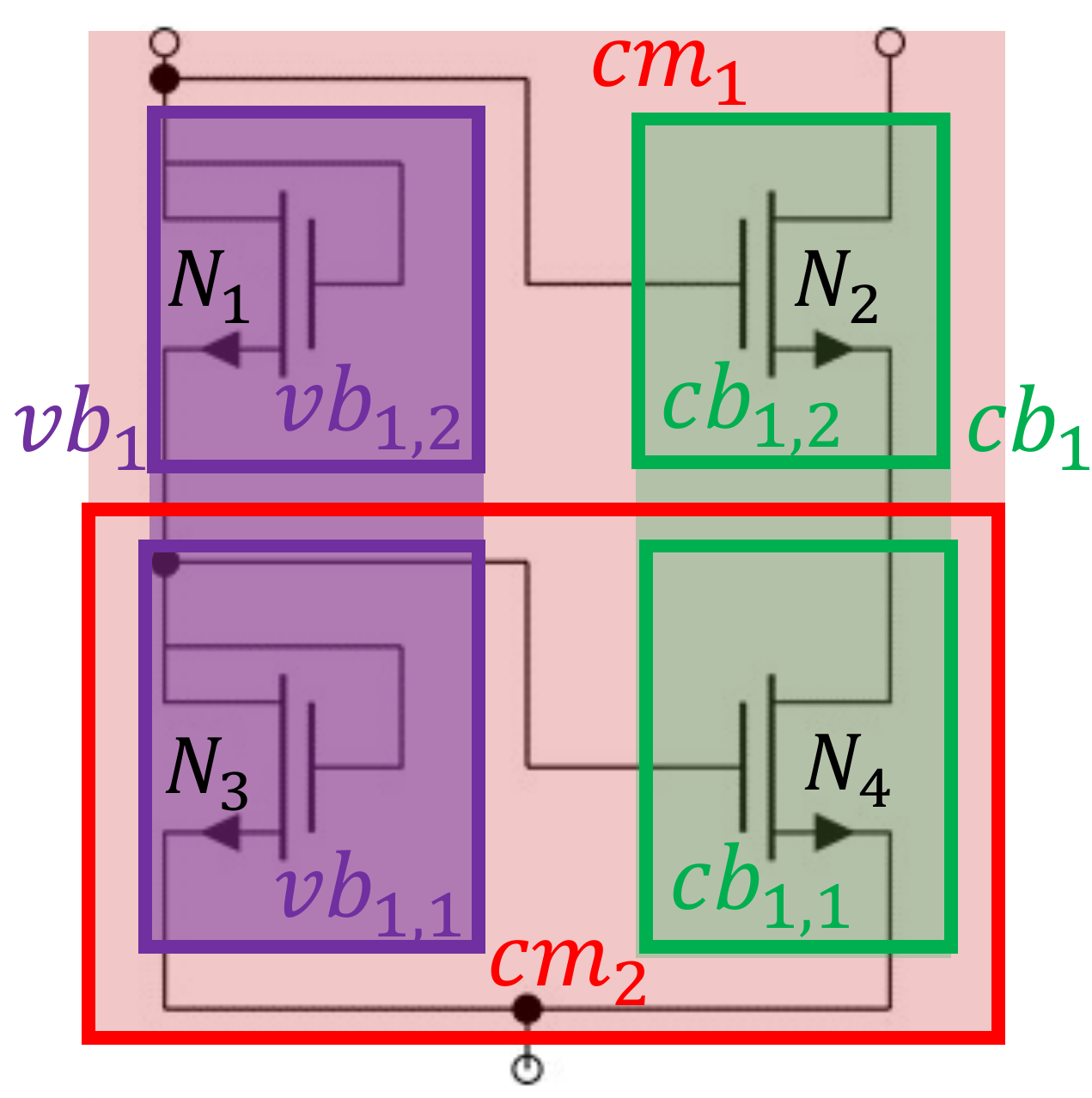}
	}%
	\qquad
	\subfloat[Cascode pairs in a differential pair with current bias]{
		\label{fig:diffStageWithCurrentMirror}
		\includegraphics[scale = 0.3]{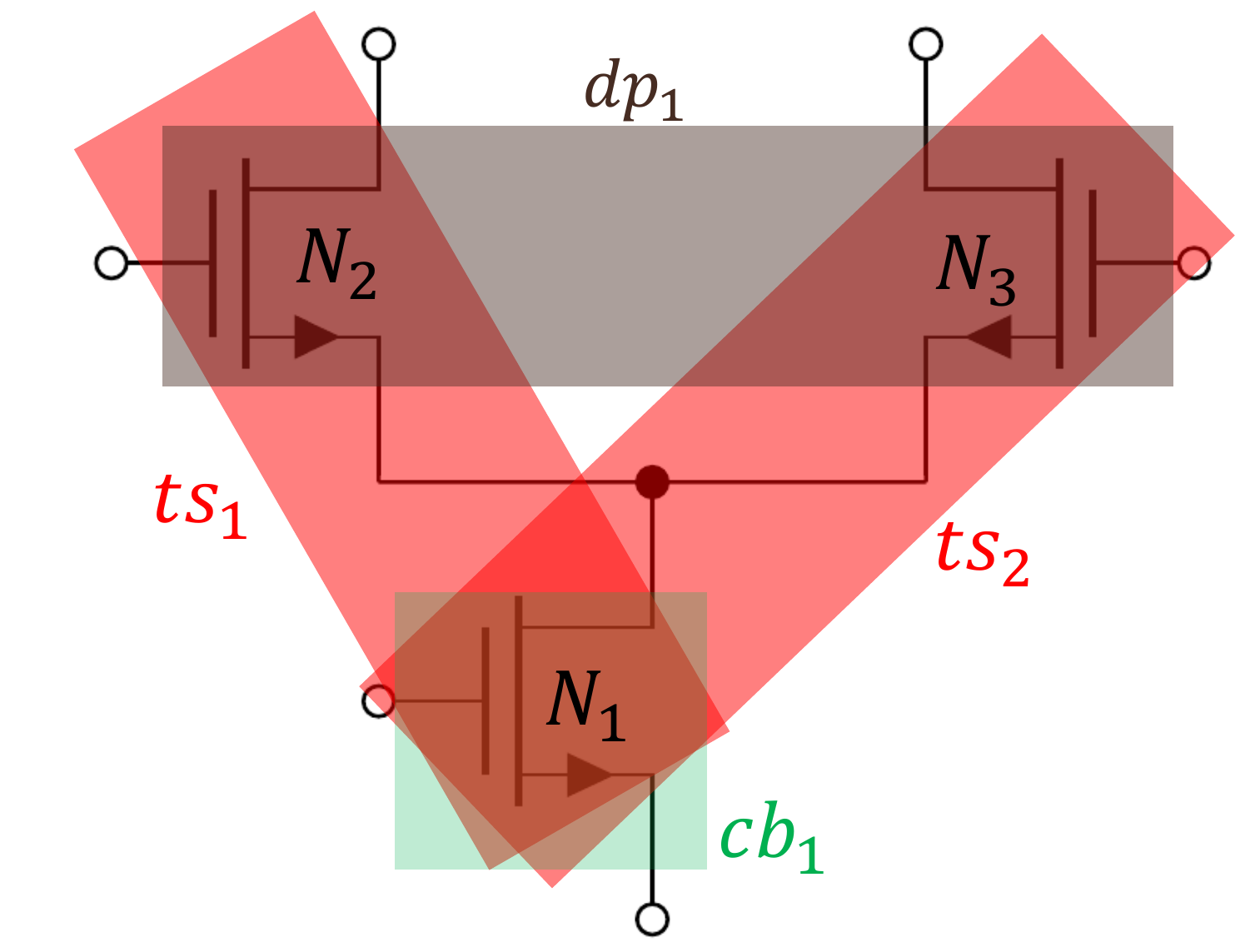}
	}%
	\qquad
	\caption{Irrelevant (a) and false (b) multiple assignments}
	\label{fig:irrelevantOrFalseAssigments}
\end{figure}

{\em Irrelevant multiple assignments}  are circuitry-wise correct but do not obtain any additional information to the functional behavior as the functional blocks the transistor is in are of the same type. 
An example is a simple current mirror in a cascode current mirror (Fig. \ref{fig:cascodeCurrentMirrorWithSCM}). The simple current mirror does not obtain any additional information to the cascode current mirror.
To avoid such irrelevant multiple assignments of transistors following rule is used: 
\begin{equation}\label{eq:irrelevantRule}
\begin{split}
&\forall_{x_k \in \mathcal{X}_{min}} [ \exists_{x_l \in \mathcal{X}} ( x_k \subset x_l  \wedge x_k.type = x_l.type)]\Leftrightarrow ~ x_k \text{is irrelevant}
\end{split}	
\end{equation} 
$\mathcal{X}_{min}$ contains  all functional blocks which are potential irrelevant. These are all current mirrors, voltage and current biases. 

{\em False multiple assignments} are circuitry-wise incorrect, e.g., two transistor stacks in a differential pair with current bias (Fig \ref{fig:inverter}). With suitable transistors connected at the drain of the differential pair, these false transistor stacks might form analog inverters.  By suppressing the recognition of these transistor stacks, the false recognition of analog inverters is omitted.

\section{Functional Blocks on Hierarchy Level 3}\label{sec:FunctionalBlockLevel3}
On HL 3  are the functional block types that form the amplification stages (HL 4) of an op-amp, i.e., transconductance, load and stage bias. 

\subsection{Transconductance}\label{sec:transconductance}

{\em Function: }
A transconductance converts a voltage potential applied at its gates into an (amplified) current.

\begin{figure}[t]\centering
	\includegraphics[width=0.99\linewidth]{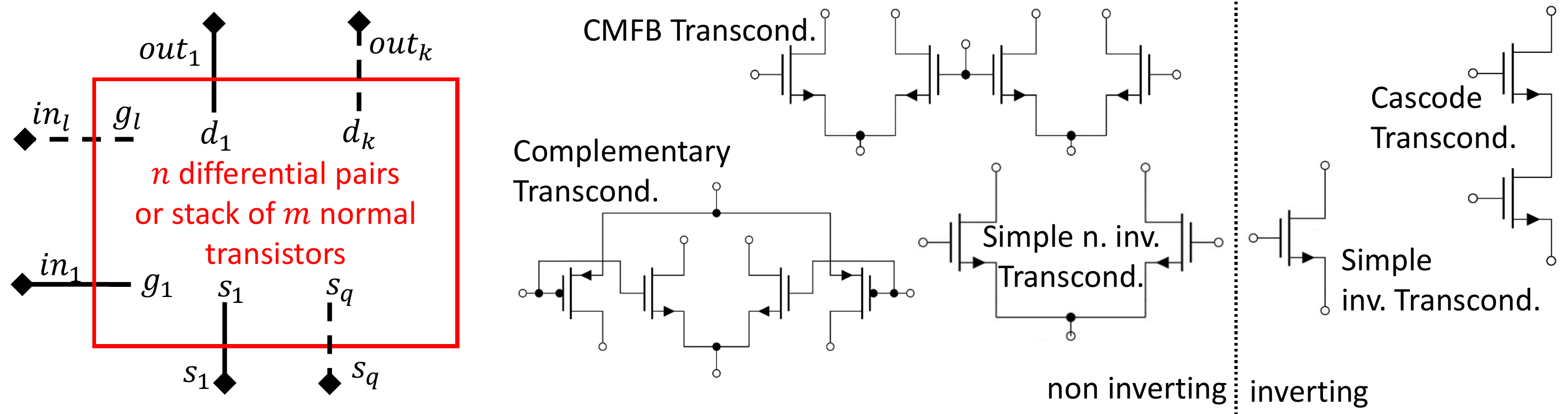}
	\caption{Transconductance (Transcond.) and examples
		(dashed lines: optional)
	}\label{fig:transconductance}
\end{figure}

{\em Structure:}
Fig. \ref{fig:transconductance} shows the structural definition of a 
transconductance $tc_k$ as well as examples.
Two different types of transconductance exist, non-inverting and inverting.

A {\em non-inverting transconductance $tc_{ninv}$} consists  of 1 or 2 differential pairs.
We further define three different types having three different structures:

A {\em simple transconductance $tc_s$} is a transconductance consisting of one differential pair, having no gate connection to any other differential pair.
\begin{equation}\label{eq:g1}
\begin{split}
x_k = &\{dp_k\} \wedge \not \exists_{{dp}_l} ~ (dp_k.g_y|_{y=1,2} \leftrightarrow dp_l.g_z|_{z=1,2}) \Leftrightarrow ~~ x_k.type = tc_s 
\end{split}
\end{equation}

A {\em complementary Transconductance $tc_c$} consists of two differential pairs with opposite doping and connected at both gates
\begin{equation}\label{eq:gc}
\begin{split}
x_k = &\{dp_{k,1}, dp_{k,2}\}
\wedge (dp_{k,1}.g_l \leftrightarrow dp_{k,2}.g_l)|_{l=1,2}   \wedge  dp_{k,1}.\Phi \neq dp_{k,2}.\Phi \\
&\Leftrightarrow  x_k.type = tc_c
\end{split}
\end{equation}

A {\em common-mode feedback (CMFB) transconductance $tc_{CMFB}$} consists of two differential pairs with equal doping connected at one of the two gates:	
\begin{equation}\label{eq:gCMFB}
\begin{split}
x_k = &\{dp_{k,1}, dp_{k,2}\}  \wedge
\exists!_{dp_{k,1}.g_m , dp_{k,2}.g_n} [dp_{k,1}.g_m \leftrightarrow dp_{k,2}.g_n] \\
& \wedge  dp_{k,1}.\Phi = dp_{k,2}.\Phi
\Leftrightarrow x_{k}.type = tc_{CMFB}
\end{split}
\end{equation}

A {\em inverting transconductance $tc_{inv}$} consists of a transistor stack $ts_k$ of $m$ normal transistors, with the source $ts_k.s$ connected to a supply voltage rail. No gate-gate and gate-drain connection of the transistors in the stack is allowed. 
The gate of the bottom transistor in the stack $ts_k.g_1$ is connected to the output of another transconductance $tc_v.out_p$ or to the output of a load $l_w.out_q$. The drain of the stack $ts_k.d$ is connected to the output of a stage bias $b_{s,y}.out_z$.
\begin{equation}\label{eq:ginv}
\begin{split}
x_k =& \{ts_k\} \wedge ts_k \subset \mathcal{X}_{nt} 
\wedge \text{net}(ts_{k}.s) \in \mathcal{N}_{supply}  \forall_{nt_i,nt_j \in ts_{k}} (nt_i.g/d \nleftrightarrow nt_j.g/d)\\
& \wedge \big[\exists_{tc_v} [tc_v.out_p \leftrightarrow ts_k.g_1 ] \vee  \exists_{l_w} [l_w.out_q \leftrightarrow ts_k.g_1 ] \big] \wedge \exists_{b_{s,y}} [b_y.out_z\leftrightarrow  ts_{k}.d]\\
&\Leftrightarrow  x_k.type = tc_{inv} 
\end{split}
\end{equation}
The definition of the load is given in the next section. The definition of the stage bias is given in Sec.~\ref{sec:bias}.

\subsection{Load}\label{sec:load}

\begin{figure} \centering
	\subfloat[A load functional block consists of one or two load parts]{
		\label{fig:loadSurblock}
		\includegraphics[scale=0.2]{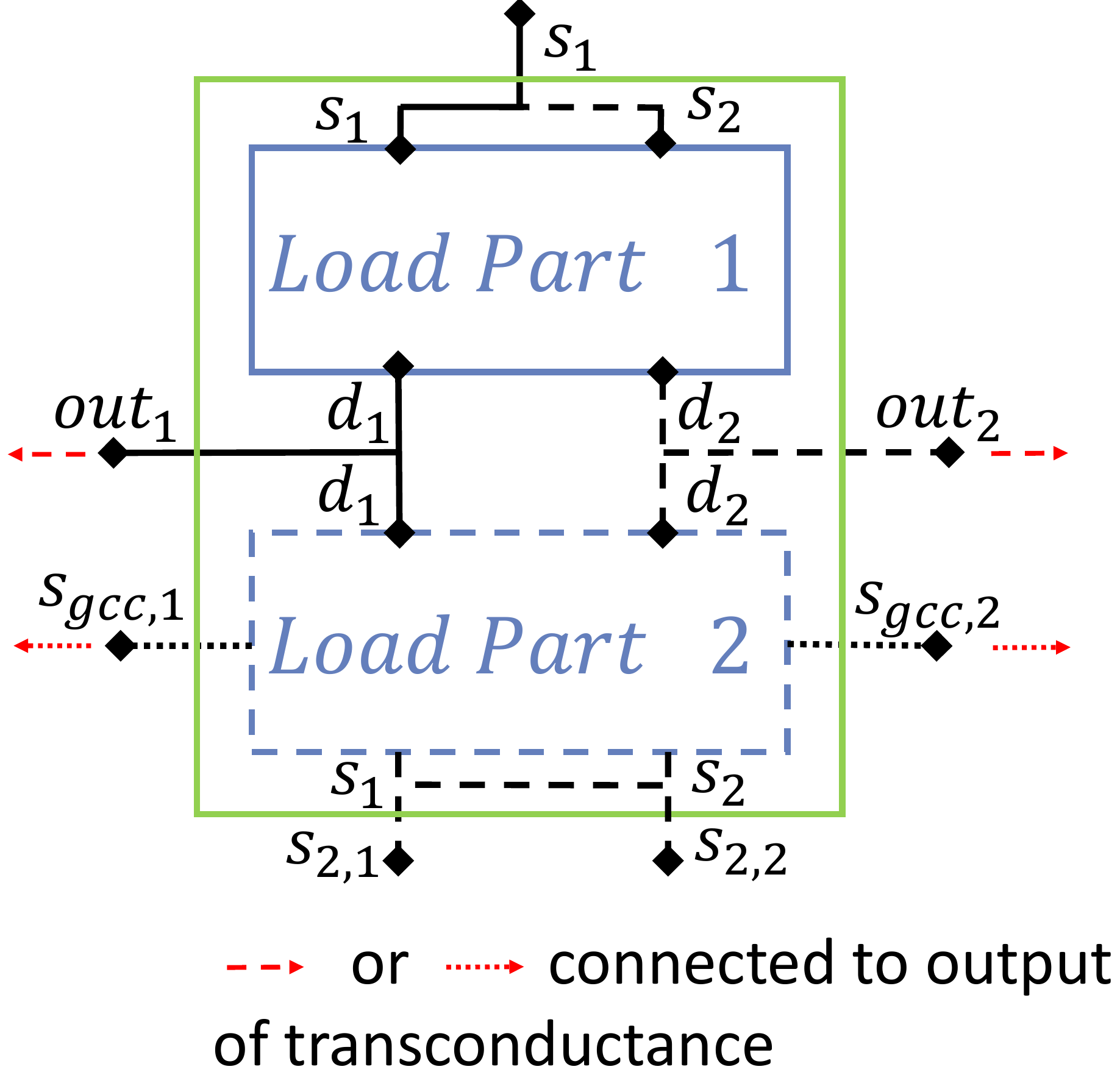}
	}%
	\qquad
	\subfloat[Load part]{
		\label{fig:loadSubblock}
		\includegraphics[scale=0.35]{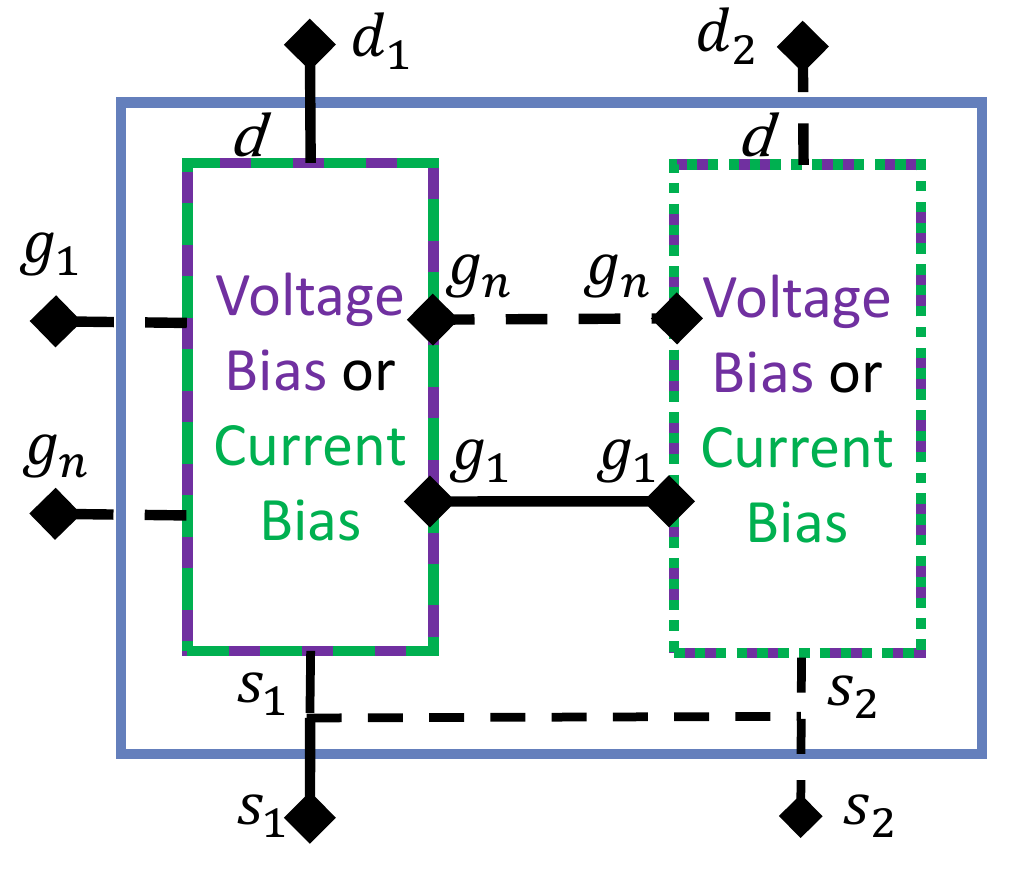}
	} \qquad
	\caption{Load
	(dashed lines: optional)
	}
	\label{fig:load}
\end{figure}
{\em Function: }
A load converts a current into a voltage. It influences the gain generated by the connected transconductance. 

{\em Structure:}
Fig. \ref{fig:load} shows the structure of a load $l_k$. A load $l_k$ consists of  one or two load parts $l_{p,k,l}$ of type $l_p$. The parts are connected at their drains and have different substrate types. If a gate connected couple $gcc_j$ is in one of the load parts, its sources $gcc_j.s_y$ are connected to the output of  a transconductance $tc_{w}$ of type $tc_{ninv}$. Otherwise, if no gate connected-couple is in one of the load parts, the outputs  of the load $l_k.out_v$ are connected to $tc_{w}.out_v$. 
\begin{equation}\label{eq:l}
\begin{split}
&x_k = \{x_ {k,1}, .., x_{k,n} |n = |x_k|\wedge n \leq 2\}
\wedge x_k \subseteq \mathcal{X}_{l_p} \wedge  \big[n = 2 \rightarrow
 [x_{k,1}.\Phi \neq x_{k,2}.\Phi \\ 
 &\wedge (x_{k,1}.d_l \leftrightarrow x_{k,2}.d_l)|_{l =1,2} ]\big]   \wedge  \bigl[ \exists_{gcc_j\in  x_k } [(gcc_j.s_y \leftrightarrow tc_z.out_y)|_{y=1,2,tc_z.type= tc_{ninv}} ]  \\
& \oplus (x_{k}.out_v \leftrightarrow tc_w.out_v )|_{v \leq |x_{k,1}|, ~tc_w.type = tc_{ninv} }  \bigr] \Leftrightarrow  x_k.type = l 
\end{split}
\end{equation}

A single load part $l_{p,k,l}$ consists of one or two transistor stacks, which have  gate-gate connections. A transistor stack in a load part is either of type voltage or current bias. If no gate connected-couple is in the load part, the sources must be connected. The doping of the transistor stacks is equal.
\begin{equation}\label{eq:lp}
\begin{split}
&x_k = \{x_ {k,1}, .., x_{k,n} |n = |x_k| \wedge n \leq 2\} 
\wedge x_{k} \subset (\mathcal{X}_{vb} \cup \mathcal{X}_{cb}) \wedge (x_{k,1}.g_l  \\
&\leftrightarrow x_{k,n}.g_l)|_{l\leq |x_{k,1}|}\wedge \not \exists_{gcc_l \in x_k} [ x_{k,1}.s\leftrightarrow x_{k,n}.s] \wedge  x_{k,1}.\Phi = x_{k,n}.\Phi  \Leftrightarrow  x_k.type = l_p
\end{split}
\end{equation}

\subsection{Stage Bias}\label{sec:bias}

{\em Function: }
A stage bias $b_s$ supplies a transconductance  with  defined currents.

\begin{figure} \centering
 	\subfloat[Bias with voltage output]{
 		\label{fig:biasVoltageBias}
 		\includegraphics[scale=0.35]{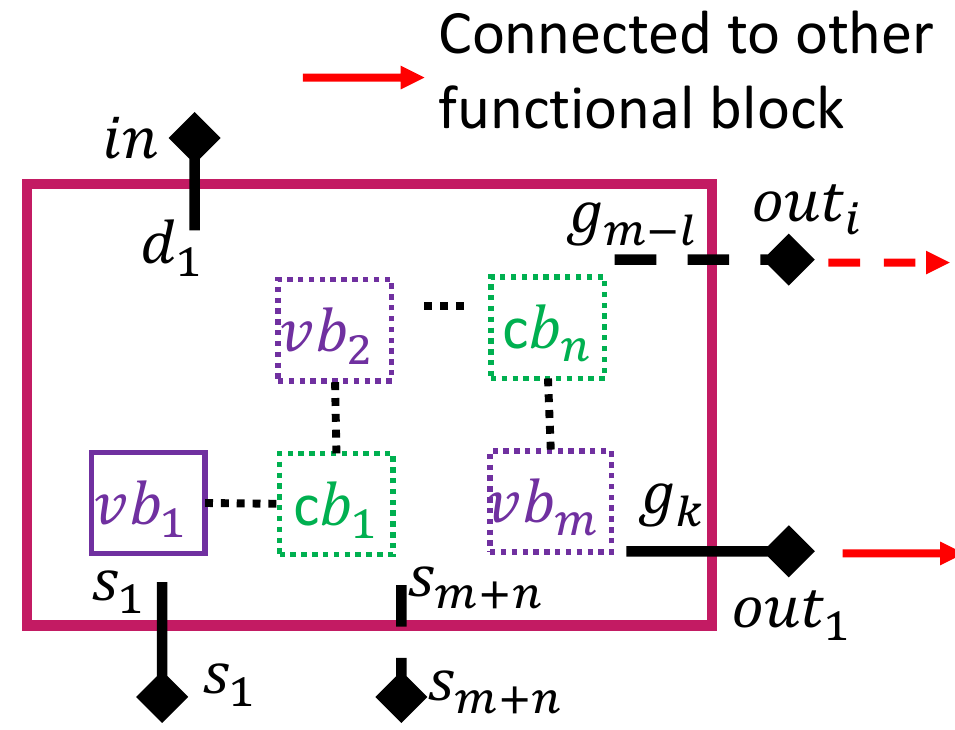}
 	}%
 	\qquad
 	\subfloat[Bias with current output]{
 		\label{fig:biasCurrentBias}
 		\includegraphics[scale=0.35]{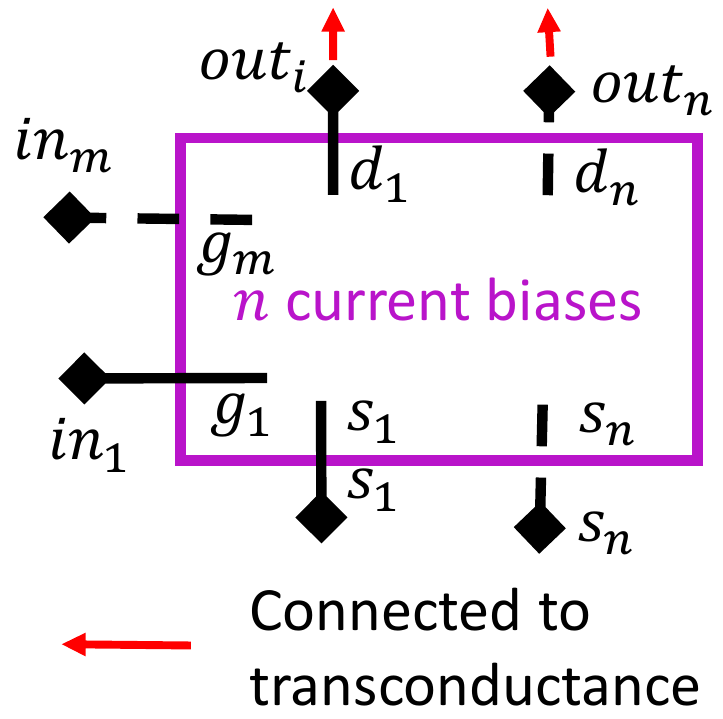}
 	} \qquad
 	\caption{Different types  of biases
 		(dashed lines: optional)
 	}
 	\label{fig:bias}
 \end{figure}

{\em Structure:}
A stage bias $b_s$ is a subtype of the type bias $b$ (Fig. \ref{fig:bias}).
 Two different types exist.
 
A {\em bias with voltage output $b_v$} consists of $m$ voltage and $n$ current biases with $m>n$. Its input $b_{v,k}.in$ is the drain of a voltage bias. The outputs $b_{v,k}.out_1, .., b_{v,k}.out_i$ are  gates of voltage biases, which are connected to gates of other functional blocks. All current biases in $b_{v,k}$ must have a drain-drain connection to a voltage bias with opposite doping and a gate-gate connection to a voltage bias with equal doping in $x_k$.  
\begin{equation}\label{eq:bv}
\begin{split}
&x_k = \{x_ {k,1}, .., x_{k,l} |l = |x_k|\} 
\wedge \{x_{k,1}, .. , x_{k,m}\} \subset \mathcal{X}_{vb}  \wedge (l \neq m \rightarrow \\
&\{x_{k,l-m},.., x_{k,l}\} \subset \mathcal{X}_{cb}  )\wedge \exists_{vb_{k,i} \in x_k} [ vb_{k,i}.g_j  \leftrightarrow x_z.g_y|_{x_z \in \mathcal{X}\setminus x_k}]\wedge\forall_{cb_{k,q} \in x_k} [ cb_{k,q}.d \\
&  \leftrightarrow vb_{k,v}.d \wedge {vb_{k,v} \in x_k} \wedge cb_{k,q}.\Phi \neq vb_{k,v}.\Phi 
\wedge (cb_{k,q}.g_s  \leftrightarrow vb_{k,w}.g.s)|_{s\leq |cb_{k,q}|} \\
&\wedge {vb_{k,w} \in x_k} \wedge cb_{k,q}.\Phi = vb_{k,w}.\Phi ]  \Leftrightarrow  x_k.type = b_v
\end{split}
\end{equation}
A bias with voltage output $b_{v,k}$ is a stage bias, iff it consists of exactly one voltage bias $vb_k$, which is connected with its drain $vb_k.d$ to the output of a transconductance of type inverting:
\begin{equation}\label{eq:bvs}
\begin{split}
&b_{v,k} = \{vb_k\} \wedge  \exists_{tc_i} [tc_i.type = tc_{inv} \wedge vb_k.d \leftrightarrow tc_i.out]
  \Leftrightarrow  b_{v,k}.type = b_s
\end{split}
\end{equation}
In all other cases, if a bias with voltage out consists of more than one voltage bias or if it is not connected with its  input pin to a transconductance, it is the circuit bias $b_O$ (Sec. \ref{sec:circuitBias})

A {\em bias with current output $b_c$} consist of $n$ current biases.  Its inputs $in_1,.. in_m$ are the gates of the current biases. The outputs  $out_1, .., out_n$ are the drains of the current biases, which must be connected to the source or output of a transconductance.
\begin{equation}\label{eq:bc}
\begin{split}
&x_k = \{cb_ {k,1}, .., cb_{k,n} |n = |x_k|\} 
\wedge x_k \subset \mathcal{X}_{cb} \wedge \forall_{cb_{k,l} \in x_k} [ cb_{k,l}.d \\
& \leftrightarrow (tc_z.s|_{tc_z.type = tc_{ninv}}  \vee tc_z.out_y|_{tc_z.type = tc_{inv}})]   \Leftrightarrow  x_k.type = b_c
\end{split}
\end{equation} 
A bias with current output is always a stage bias: 
\begin{equation}\label{eq:bcs}
\begin{split}
\forall_{b_k \in \mathcal{X}_{b}} [b_k.type = b_c \rightarrow b_k.type = b_s]
\end{split}
\end{equation}

\section{Functional Blocks on Hierarchy Level 4}\label{sec:FunctionalBlockLevel4}
The functional block types on HL 4 are the amplification stage, the circuit bias, the compensation and  load capacitor.

\subsection{Amplification Stage}\label{sec:amplificationStage}

\begin{figure}\centering
	\includegraphics[width=0.75\linewidth]{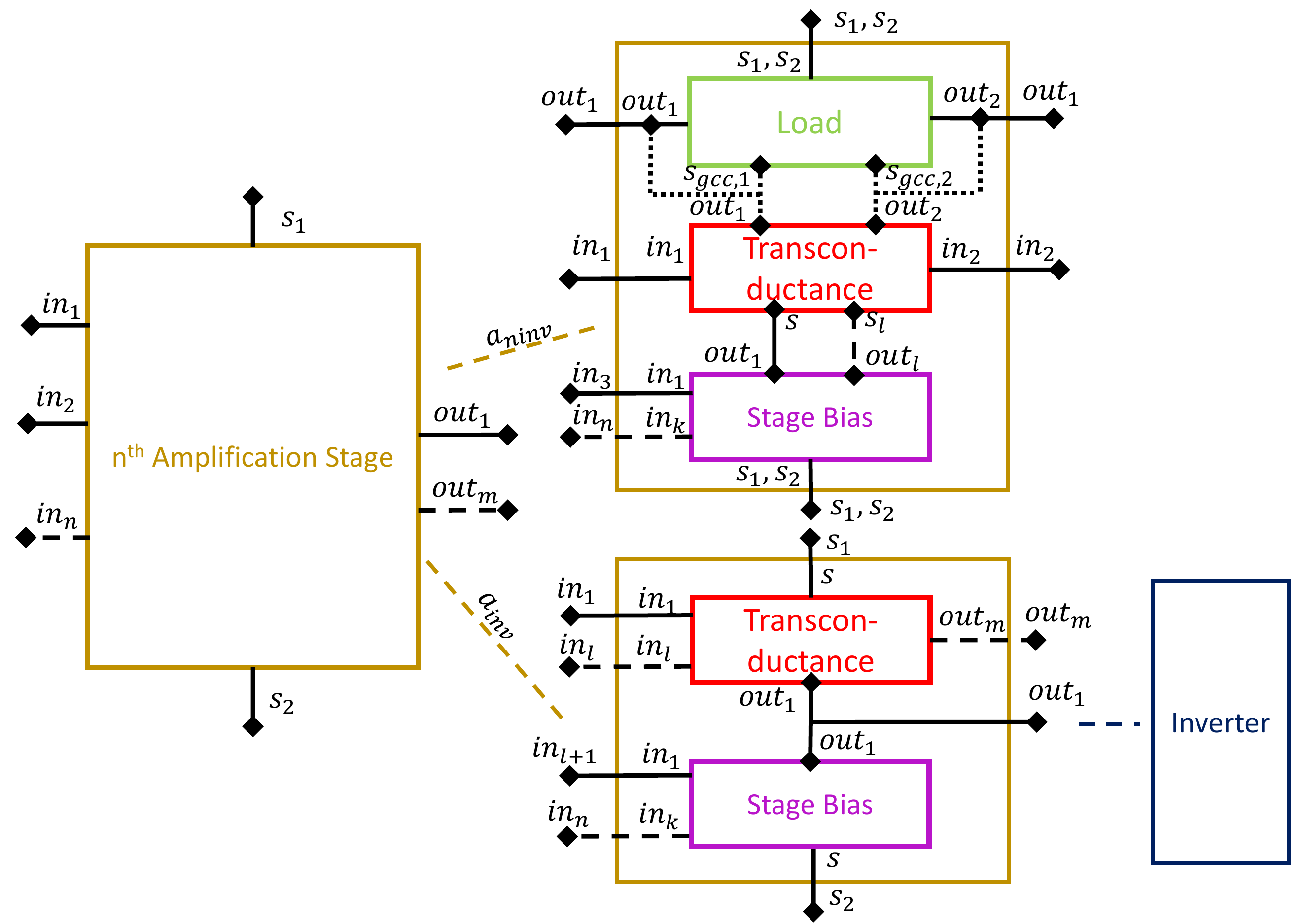}
	\caption{Amplification stage
	(dashed lines: optional)}\label{fig:functionalBlocksAmplificationStages}
\end{figure} 
{\em Function: }
Two functional types of amplification stage exist: the $n^{th}$ amplification stage of an op-amp $a_{n}$ and the common-mode feedback stage $a_{CMFB}$.
The $n^{th}$ amplification stage $a_{n}$ amplifies the input signal for the $n^{th}$ time.
A common-mode feedback (CMFB) stage $a_{CMFB}$ amplifies the output signals of a fully differential op-amp while comparing them to a reference voltage and feeds them back to the amplifier.

{\em Structure:}
Fig. \ref{fig:functionalBlocksAmplificationStages} shows the general composition of the amplification stages of an op-amp.  Every amplification stage has at least two inputs and one output. It has a source connection to both supply voltage rail.
Two structural types of amplification stages exist, non-inverting and inverting amplification stages. Non-inverting amplification stages are the first stage of an op-amp and the CMFB stage. Inverting stages form the further stages of an op-amp ($n \geq 2$). In the following, both types are described in detail.

A {\em non-inverting amplification stage $a_{ninv}$} consists of a transconductance $tc_k$, a load $l_k$ and a stage bias $b_{s,k}$. The transconductance must be of type non-inverting and the stage bias must have a current output connected to the sources of the transconductance. The load outputs or the sources of a gate connected couple in the load must be connected to the outputs of  the transconductance.
\begin{equation}\label{eq:aninv}
\begin{split}
&x_{k} = \{tc_ {k},l_{k}, b_{s,k}\} 
\wedge tc_{k}.type = tc_{ninv} \wedge b_{s,k}.type = b_c \wedge \big[ \exists_{gcc_i \in l_{k}} [(gcc_i.s_j \\
& \leftrightarrow tc_{k}.out_j)|_{j=1,2}] \oplus \forall_{l_{k}.out_v \in P_{l_{k}}} [l_{k}.out_v \leftrightarrow tc_{k}.out_w] \big]  \\
&\wedge (tc_{k}.s_l \leftrightarrow b_{s,k}.out_l)|_{l \leq |tc_k|} \Leftrightarrow  x_k.type = a_{ninv} 
\end{split}
\end{equation}

Non-inverting amplification stages are classified into three types: simple first stage $a_s$, complementary first stage $a_c$, and CMFB stage $a_{CMFB}$. The types differ in the transconductance types and the doping characteristics of its  functional blocks.

A {\em simple first stage $a_s$} has a transconductance of type $tc_s$. The transconductance and the stage bias have the same substrate type, while the load has either the opposite doping or mixed doping. 
\begin{equation}\label{eq:a1}
\begin{split}
&a_{ninv} = \{tc_{k},l_{k}, b_{s,k}\} 
\wedge tc_{k}.type  = tc_s  \wedge (tc_k.\Phi = b_{s,k}.\Phi) \in {\{\Phi_n, \Phi_p \}} \\
&\wedge tc_k.\Phi \neq l_k.\Phi\Leftrightarrow  a_{ninv}.type = a_{s} 
\end{split}
\end{equation}

A {\em complementary first stage $a_c$} has a transconductance of type $tc_c$. Transconductance, load and stage bias have all mixed doping. 
\begin{equation}\label{eq:ac}
\begin{split}
&a_{ninv} = \{tc_{k},l_{k}, b_{s,k}\} 
\wedge tc_{k}.type  = tc_c  \wedge tc_k.\Phi = b_{s,k}.\Phi = l_k.\Phi = \Phi_u  \\
&\Leftrightarrow  a_{ninv}.type = a_{c} 
\end{split}
\end{equation}

A {\em common-mode feedback stage $a_{CMFB}$} has a transconductance of type $tc_{CMFB}$. Transconductance and stage bias have the same substrate type, while the load has opposite doping. 
\begin{equation}\label{eq:aCMFB}
\begin{split}
&a_{ninv} = \{tc_{k},l_{k}, b_{s,k}\} 
\wedge tc_{k}.type  = tc_{CMFB}  \wedge (tc_k.\Phi = b_{s,k}.\Phi) \in \{\Phi_n, \Phi_p \}   \\
& \wedge tc_k.\Phi \neq l_k.\Phi|_{l_k.\Phi \in \{\Phi_n, \Phi_p \}} \Leftrightarrow  a_{ninv}.type = a_{CMFB} 
\end{split}
\end{equation}

A {\em inverting amplification stage $a_{inv}$} consists of a transconductance $tc_{k}$ and a stage bias $b_{s,k}$ with  opposite doping connected at their outputs. The transconductance must be of type inverting. The stage bias consists either of one voltage bias or one  current bias.
\begin{equation}\label{eq:ainv}
\begin{split}
&a_{k} = \{tc_{k},b_{s,k}\} 
\wedge tc_{k}.type = tc_{inv} \wedge b_{s,k}.type \in \{b_c, b_v\}  \wedge |b_{s,k}|= 1 
\\
\wedge & tc_k.out_1 \leftrightarrow b_{s,k}.out_1 \wedge tc_k.\Phi \neq b_{s,k}.\Phi  \Leftrightarrow  a_k.type = a_{inv} 
\end{split}
\end{equation}
Note, that all inverting stages can occur twice in an op-amp, e.g., two second stages. It appears when the op-amp is fully differential or a symmetrical op-amp (Fig. \ref{fig:symmetricalOpAmpWithSecondStage}, with $a_{2,1}, a_{2,2}$).

Differentiating between an inverting stage with a stage bias with current  output and an inverting stage with a stage bias with voltage output allows to give a more precise definition of inverting amplification stages:

A {\em inverting stage with stage bias with current output $a_{inv_c}$} must also fulfill the type definition of an analog inverter to be a valid  inverting amplification stage:
\begin{equation}\label{eq:ainvc}
\begin{split}
&a_{inv,k} = \{tc_{inv, k},b_{s,k} \} 
\wedge b_{s,k}.type = b_{c} \wedge a_{inv,k}.type = inv \Leftrightarrow  a_{inv,k}.type = a_{inv_c} 
\end{split}
\end{equation}
Note, that not all analog inverters are inverting stages, as the constraints for a transconductance \eqref{eq:ginv} and stage bias \eqref{eq:bc} must hold.

A {\em inverting stage with stage  bias with voltage output $a_{inv_v}$} only occurs in symmetrical OTAs, which are op-amps with a characteristic first stage and two second stages (Fig. \ref{fig:symmetricalOpAmpWithSecondStage}). The first stage $a_1$ must be a simple first stage $a_s$ having a load consisting of two voltage biases $vb_{l,1},vb_{l,2}$  with same doping. Both voltage biases must have a gate-gate connection to a transconductance in a second stage of type $a_{inv}$. One second stage $a_z$ must be of type $a_{inv_c}$. The other second stage $a_y$ must have a voltage bias as stage bias, which has a gate-gate connection to the stage bias of $a_z$:
\begin{equation}\label{eq:ainvv}
\begin{split}
&a_{inv,k} = \{tc_{inv, k},b_{s,k} \} 
\wedge b_{s,k}.type = b_{v} \wedge \exists_{a_1} \Bigl[ a_1.type = a_s \\
&  \wedge \exists_{l_1 \in a_1} \big[ |l_1| = 1 \wedge \exists_{l_{1,1} \in l_1} [ \{vb_{l,1}, vb_{l,2}\} \subseteq l_{1,1} \wedge ( vb_{l,m}.g_1 \leftrightarrow tc_{inv,k}.in_1  \\
&\wedge vb_{l,n}.g_1 \leftrightarrow a_i.in_1 )|_{(m=1 \wedge n =2) \vee (m=2 \wedge n =1)} \wedge a_i.type = a_{inv_c}\\
&  \wedge  (b_{s,k}.out_v \leftrightarrow b_i.in_v|_{b_i \in a_i})]\big] \Bigr]  \Leftrightarrow  a_{inv,k}.type = a_{inv_v} 
\end{split}
\end{equation}	  
 
\subsection{Circuit Bias}\label{sec:circuitBias}
{\em Function:}
The circuit bias $b_O$ supplies all functional blocks of type amplification stage $a$ with voltages specified by $b_O$.

{\em Structure:}
The circuit bias $b_O$ has the structure of a bias with voltage output \eqref{eq:bv} (Fig. \ref{fig:biasVoltageBias}). It contains all current biases and voltage biases not part of an amplification stage.
\begin{equation}\label{eq:bo}
\begin{split}
	x_k =& \{x_{k,1}, .. , x_{k,n}| n = |x_{k}|\} \wedge x_k = [(\mathcal{X}_{vb} \cup \mathcal{X}_{cb})\setminus\mathcal{X}_a] \wedge x_k.type = b_v\\
	& \Leftrightarrow x_k.type = b_O
	\end{split}
\end{equation}

\subsection{Compensation Capacitor}\label{sec:compCap}
{\em Function: }
A compensation capacitor $c_{C,k}$ increases the stability of an op-amp.

{\em Structure:}
A compensation capacitor $c_{C,k}$ is connected between the outputs of two different amplification stages $a_j,a_v$:
\begin{equation}\label{eq:cC}
\begin{split}
x_k = & \{cap_k\} \wedge [cap_k.p_1 \leftrightarrow a_j.out_i 
\wedge cap_k.p_2 \leftrightarrow a_v.out_w]|_{a_j \neq a_v} 
\Leftrightarrow ~ x_k.type = c_C
\end{split}
\end{equation}

\subsection{Load Capacitor}
{\em Function: }
The load capacitor $c_{L,k}$ represents the capacitive load the op-amp is able to drive in its application.

{\em Structure:}
A load capacitor $c_{L,k}$ is connected between an output of the highest ($n$-th) amplification stage $a_{n}$ and ground:
\begin{equation} \label{eq:cL}
\begin{split}
x_k = & \{cap_k\} \wedge cap_k.p_1 \leftrightarrow a_{n}.out_j \wedge \text{net}(cap_k.p_2) = GND
\Leftrightarrow ~ x_k.type = c_L
\end{split}
\end{equation}

\section{Functional Block Analysis}\label{sec:functionalBlockAnalysis}

\begin{figure}\centering
	\includegraphics[width=0.55\linewidth]{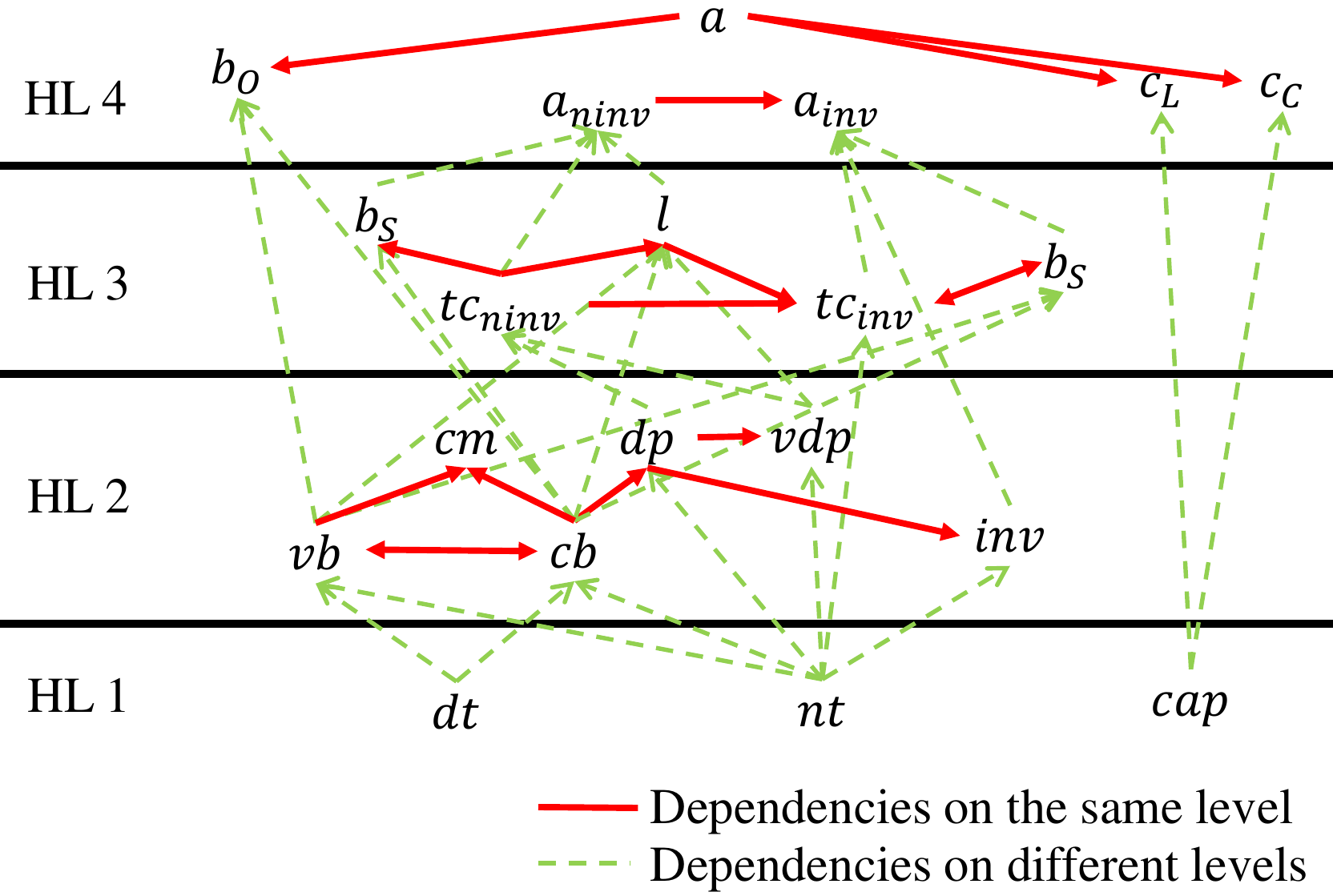}
	\caption{Dependency graph of the functional blocks in op-amps (see Fig.~\ref{fig:functionalBlockPyramid})}\label{fig:dependencyGraph}
\end{figure}

The functional block analysis recognizes all functional blocks in an op-amp netlist based on the definitions given in Sec. \ref{sec:DeviceLevelFunctionalBlock} - \ref{sec:FunctionalBlockLevel4}.
Fig. \ref{fig:dependencyGraph} shows the dependency graph that arises from these definitions. Note, that bidirectional dependencies exist. Therefore complex algorithms are required for the functional block analysis.

\subsection{Hierarchy Level 1}

On hierarchy level 1, the definitions of the functional block types are independent of each other. Therefore,  the definitions given in Sec. \ref{sec:DeviceLevelFunctionalBlock} can be used for recognition without any restriction to their order.

\subsection{Hierarchy Level 2}

\begin{algorithm} [t]{\small
		\caption{\small Recognition of functional block types on hierarchy level 2}
		\label{algo:HierarchlevelTwo}
		\begin{algorithmic}[1]
			\REQUIRE  $\mathcal{X}$ = $\mathcal{X}_{nt} \cup \mathcal{X}_{dt} \cup \mathcal{X}_{dt}$
			\STATE $\mathcal{X}_{vb} := \{ \}$ //At the beginning the set of voltage biases is empty
			\STATE $\mathcal{X}_{cb} := \{ \}$ //At the beginning the set of current biases is empty
			\STATE $\mathcal{X}_{ts_n}:=$findAllTransistorStacksNMOS($\mathcal{X}$)  //Definition \eqref{eq:ts}\label{line:FindAllTransistorStacksNMOS}
			\STATE $\mathcal{X}_{ts_p}:=$findAllTransistorStacksPMOS($\mathcal{X}$) //Definition \eqref{eq:ts}\label{line:FindAllTransistorStacksPMOS}
			\REPEAT
			\FORALL {$ts_k \in [\mathcal{X}_{ts_j}|_{j \in \{n,p\}}\setminus (\mathcal{X}_{vb} \cup \mathcal{X}_{cb})] $}
			\IF {$\exists_{ts_l \in \mathcal{X}_{ts_j}} \big[ ts_k.d \leftrightarrow ts_l.g_m \wedge  \forall_{ts_{k}.g_v} 
				[ts_{k}.g_v \leftrightarrow ts_l.g_w  ]\big] $} \label{line:VoltageBiasCheck}
			\IF{$ \forall_{ts_l.g_m} [ts_{l}.g_m \leftrightarrow(ts_k.g_m \vee ts_k.d)] \wedge$\\ $\not\exists_{ts_i \in \mathcal{X}_{ts_j}} [  ts_i.g_n \leftrightarrow ts_l.d]$}\label{line:CurrentBiasCheck}
			\STATE $\mathcal{X}_{vb} := \mathcal{X}_{vb} \cup \{ts_k\}$
			\STATE $\mathcal{X}_{cb} := \mathcal{X}_{cb} \cup \{ts_l\}$
			\ELSIF {$ts_l \in \mathcal{X}_{cb}$} \label{line:IdentifiedCurrentBias?}
			\STATE $\mathcal{X}_{vb} := \mathcal{X}_{vb} \cup \{ts_k\}$
			\ENDIF
			\ENDIF
			\ENDFOR
			\UNTIL{no new voltage or current bias was found}
			\STATE $\mathcal{X}_{cm} :=$ findAllCurrentMirrors($\mathcal{X}_{vb}, \mathcal{X}_{cb}$) //Definition \eqref{eq:cm}
			\STATE $\mathcal{X}_{dp}, \mathcal{X}_{vdp} :=$ findAllDifferentialPairs($\mathcal{X}_{nt}, \mathcal{X}_{cb}$) //Definitions \eqref{eq:dp}, \eqref{eq:vdp}
			\STATE $\mathcal{X}_{inv} :=$ findAllAnalogInverters($\mathcal{X}_{ts_n}, \mathcal{X}_{ts_p}$) //Definition \eqref{eq:inv}
			\STATE $\mathcal{X} := \mathcal{X} \cup \mathcal{X}_{vb} \cup \mathcal{X}_{cb} \cup \mathcal{X}_{cm} \cup \mathcal{X}_{dp} \cup\mathcal{X}_{vdp} \cup \mathcal{X}_{inv}$ 
			\STATE $\mathcal{X} :=$ deleteIrrelevantStructures($\mathcal{X}$) //Definition \eqref{eq:irrelevantRule}
			\RETURN $\mathcal{X}$
	\end{algorithmic}}
\end{algorithm}

On  level 2, every recognition of a functional block type depends on another functional block on the same  level \eqref{eq:vb} - \eqref{eq:inv}. 
Voltage bias and current bias are at the starting point of the dependency graph being bidirectionally dependent.
Alg. \ref{algo:HierarchlevelTwo} presents the recognition algorithm.

To resolve all dependencies, we propose to start with the recognition of transistor stacks \eqref{eq:ts}, which are auxiliary blocks independent of all functional block types of level 2. On basis of the recognized transistor stacks, the current and voltage biases are recognized. For every substrate type, a set of transistor stacks $\mathcal{X}_{ts_j}|_{j\in \{n,p\}}$ is created (Line \ref{line:FindAllTransistorStacksNMOS}, \ref{line:FindAllTransistorStacksPMOS}). The algorithm iterates over both sets. It  searches for two transistor stacks $ts_k, ts_l$, which are connected at their gates and also have a drain-gate connection $ts_k.d \leftrightarrow ts_l.g_m$ (Line \ref{line:VoltageBiasCheck}). This corresponds to the definition of a voltage bias \eqref{eq:vb}, taking $ts_k$ as voltage bias  connected to a current bias $ts_l$.
It is checked if $ts_l$ has all the  gate-drain and gate-gate  connections to $ts_k$ needed for a current bias (Line \ref{line:CurrentBiasCheck}), and if it is not connected with its drain to any gate of a transistor stack in $\mathcal{X}_{ts_j}$. This corresponds to the definition of a current bias \eqref{eq:cb}.
If this is the case, the  $ts_k$ and  $ts_l$ are {\em primary voltage and current biases}, i.e., they  have all gate-gate and gate-drain connection to each other.
Another case is that $ts_l$ is an already identified current bias (Line \ref{line:IdentifiedCurrentBias?}). In this case, the primary voltage and current bias were already recognized and a {\em secondary voltage bias} $ts_k$ is recognized, e.g. a voltage bias which biases the uppermost gate of a wide-swing cascode current mirror.
Note, that the last part of the voltage bias definition \eqref{eq:vb} is not checked  as this is always ensured by using an valid op-amp topology.

The repetitive iteration over all recognized transistor stacks is needed to ensure that all secondary voltage biases are found.
After finding all voltage and current biases. The current mirrors and differential pairs are found using their definitions.  The analog inverter is the last functional block of HL 2, which is recognized to omit their false recognition in differential pairs.
With \eqref{eq:irrelevantRule} all irrelevant functional blocks are deleted from the set of all functional blocks. 

\subsection{Hierarchy Level 3 - 4}

\begin{algorithm}  {\small
	\caption{\small Recognition of functional block types on the hierarchy levels 3-4}
	\label{algo:HierarchlevelThrreToFive}
	\begin{algorithmic}[1]
		\REQUIRE  $\mathcal{X}$
		\STATE //Searching for non-inverting stages
		\STATE  $\mathcal{X}_{tc_{ninv}}:=$findAllNonInvertingTransconductance($\mathcal{X}_{dp}, \mathcal{X}_{vdp}$) //Definitions \eqref{eq:g1}, \eqref{eq:gc}, \eqref{eq:gCMFB} 
		\STATE $\mathcal{X}_{l}:=$findAllLoads($\mathcal{X}_{tc_{ninv}}, \mathcal{X}$) //Algorithm \ref{algo:loadRecognition}, 
		\STATE $\mathcal{X}_{b_{s,ninv}}:=$findStageBiases($\mathcal{X}_{tc_{ninv}}, \mathcal{X}_{cb}$) //Definition \eqref{eq:bcs}
		\STATE $\mathcal{X}_{a_{ninv}}:=$findAllNonInvertingStages($\mathcal{X}_{tc_{ninv}}, \mathcal{X}_{l}, \mathcal{X}_{b_{s,ninv}}$) //Definitions \eqref{eq:a1}, \eqref{eq:ac}, \eqref{eq:aCMFB}
		\STATE $\mathcal{X} := \mathcal{X} \cup \mathcal{X}_{tc_{ninv}} \cup \mathcal{X}_l \cup \mathcal{X}_{b_{s,ninv}} \cup \mathcal{X}_{a_{ninv}}$
		\STATE //Searching for inverting stages
		\STATE $\mathcal{X}_{a_{inv}} = \{ \}$
		\REPEAT
		\FORALL{$inv_k \in \mathcal{X}_{inv}$}
		\IF {$\exists_{ts_{k,1} \in inv_k} [\exists_{a_i \in \mathcal{X}_a} a_i.out_j \leftrightarrow ts_{k,1}.g_1 ] \wedge \exists_{ts_{k,2} \in inv_k} (ts_{k,2}.type = cb)$}
		\STATE $tc_{inv,k} = ts_{k,1}$
		\STATE $b_{s,inv,k} = ts_{k,2}$
		\STATE $a_{inv,k} = \{tc_{inv,k}, b_{s,inv,k}\}$
		\STATE $\mathcal{X} := \mathcal{X} \cup \{tc_{inv,k}, b_{s,inv,k}, a_{inv,k}\}$
		\ENDIF 
		\IF{$|\mathcal{X}_{a_{inv}}| =1 \wedge \exists_{a_1} \big[a_1.type = a_s \wedge \exists_{l_1 \in a_1} [ |l_1| =1 \wedge \exists_{l_{1,1} \in l_1} ( \{vb_{l,1}, vb{l_2} \subseteq l_{1,1}) ] ]\big]$} \label{line:criteriaSymmetricalOpAmp}
		\IF {$\exists_{ts_i} [ts_i.g_1 \leftrightarrow (vb_{l,1}.g_1 \vee vb_{l,2}.g_1) \wedge ts_i.d \leftrightarrow vb_j.d \wedge (vb_j.g_q \leftrightarrow a_{inv,1}.g_q|_{a_{inv,1} \in \mathcal{X}_{a_{inv}}})|_{q \leq |vb_j|} ]$}
		\STATE $tc_{inv,i} = ts_{i}$
		\STATE $b_{s,inv,v} = vb_{j}$
		\STATE $a_{inv,v} = \{tc_{inv,i}, b_{s,inv,v}\}$
		\STATE $\mathcal{X} := \mathcal{X} \cup \{tc_{inv,i}, b_{s,inv,v}, a_{inv,v}\}$
		\ENDIF 
		\ENDIF 
		\ENDFOR
		\UNTIL{no new stage is found}
		\STATE $\mathcal{X}_{b_O}$ := findCircuitBias($\mathcal{X}_{cb} \cup \mathcal{X}_{vb}, \mathcal{X}_a$) //Definition \eqref{eq:bo}
		\STATE $\mathcal{X}_{c_C}$ := findCompensationCapacitors($\mathcal{X}_{cap}$,$\mathcal{X}_a$) //Definition \eqref{eq:cC}
		\STATE $\mathcal{X}_{c_L}$ := findLoadCapacitors($\mathcal{X}_{cap}$,$\mathcal{X}_a$)  //Definition \eqref{eq:cL}
		\STATE $\mathcal{X} := \mathcal{X} \cup \mathcal{X}_{b_O} \cup \mathcal{X}_{c_C} \cup \mathcal{X}_{c_L}$
		\RETURN $\mathcal{X}$
	\end{algorithmic}}
\end{algorithm}

We combine the recognition of the functional block types of HL 3 and HL 4 as this simplifies the recognition process and resolves the bidirectional dependency of the inverting transconductance $tc_{inv}$ and its stage bias $b_s$.
The algorithm (Alg.~\ref{algo:HierarchlevelThrreToFive}) starts by recognizing all non-inverting transconductances. They are the only functional blocks on HL 3 that are independent of any other functional block \eqref{eq:aninv}.

In the next step, the loads are recognized. Here we propose not to use the exact definition of the load \eqref{eq:l} but instead use the algorithm shown in Alg. \ref{algo:loadRecognition}.
The advantage of this algorithm is that it does not depend on recognized voltage or current biases as the definition \eqref{eq:lp} does. Often external voltage biases are used to bias the load. In this case, current biases as load are not recognized. Alg. \ref{algo:loadRecognition} is more general and uses transistor stacks for recognition. It first searches for the nets, to which the load parts are connected (Line \ref{line:searchingForNetBegin} - \ref{line:searchingForNetEND}). This is either the drain net of a gate connected couple $gcc_l.d$  or if no $gcc_l$ exists, the drain nets of differential pairs in $tc_{ninv,k}$.  At these nets, the algorithm searches for the transistor stacks forming the load (Line \ref{line:searchingForTransistorStacksBegin} - \ref{line:searchingForTransistorStacksEnd}). Note, that a gate connected couple if recognized, is part of a load while a differential pair is not.

After the recognition of the loads, the stage biases of non-inverting stages are recognized by definition \eqref{eq:bcs}. With the non-inverting transconductances, loads and stage biases, the non-inverting stages are recognized.

\begin{algorithm} [tb]{\small
		\caption{\small findAllLoads($\mathcal{X}_{tc_{ninv}}, \mathcal{X}$)}  \label{algo:loadRecognition}
		\begin{algorithmic}[1]
			\REQUIRE $\mathcal{X}_{tc_{ninv}}$, $\mathcal{X}$
			\STATE $\mathcal{X}_l:= \{\}$
			\FORALL{$tc_{ninv,k} \in \mathcal{X}_{tc_{ninv}}$} 
			\STATE $\mathcal{N} := \{ \}$ // Set of nets the load parts are connected to is empty
			\IF {$\exists_{dp_l \in tc_{ninv,k}} \{dp_l\} \subset vdp_l $}\label{line:searchingForNetBegin}
			\STATE $gcc_l = vdp_l \setminus \{dp_l\}$
			\STATE  $\mathcal{N} := \mathcal{N} \cup \{\text{net}(gcc_l.d_1),\text{net}( gcc_l.d_2)\}$
			\ELSE
			\STATE $\mathcal{N} := \mathcal{N} \cup \{\text{net}(tc_{inv,k}.out_1), \text{net}(tc_{inv,k}.out_2), ..\}$
			\ENDIF \label{line:searchingForNetEND}
			\STATE $l_{p,n}$:=$\{ \}$//The load part containing NMOS transistors is empty 
			\STATE $l_{p,p}$:=$ \{ \}$//The load part containing PMOS transistors is empty 
			\FORALL{${n_j} \in \mathcal{N}$} \label{line:searchingForTransistorStacksBegin}
			\IF {$\exists_{ts_m} \big[ts_m.\Phi = \Phi_n \wedge \text{net}(ts_m.d) = n_j\wedge [ts_m.s \leftrightarrow (GND \vee tc_{ninv,k}.out_z)] \big] $}
			\STATE $l_{p,n} := l_{p,n} \cup \{ts_m\}$
			\ENDIF
			\IF {$\exists_{ts_n} \big[ts_n.\Phi = \Phi_p \wedge\text{net}(ts_n.d) = n_j \wedge [ts_n.s \leftrightarrow (VDD \vee tc_{ninv,k}.out_z)] \big] $}
			\STATE $l_{p,p} := l_{p,p} \cup \{ts_n\}$
			\ENDIF
			\ENDFOR \label{line:searchingForTransistorStacksEnd}
			\STATE $\mathcal{X}_l := \mathcal{X}_l \cup \{l_{p,p}, l_{p,n}\}$
			\ENDFOR
			\RETURN $\mathcal{X}_l$
	\end{algorithmic}}
\end{algorithm}
In the next step, inverting stages are recognized.
For recognition, we use that an inverting stage is also a functional block of type analog inverter iff its stage bias is of type current bias \eqref{eq:ainvc}. This resolves the bidirectional dependency of an inverting transconductance $tc_{inv,k}$ and its bias $b_{s,inv,k}$.   
The algorithm iterates over all recognized analog inverters. For every analog inverter, it is checked, if one of its stacks $ts_{k,1}$ is connected with its first gate $ts_{k,1}.g_1$ to the output of an already recognized stage $a_i$. This corresponds to the definition of an inverting transconductance $tc_{inv,k}$ \eqref{eq:ginv}. The output of a stage is either the output of its load or the output of its transconductance (Fig. \ref{fig:transconductance}).
The other transistor stack $ts_{k,2}$ in the analog inverter must be of type current bias and thus is the stage bias \eqref{eq:bcs} of the inverting stage.

After finding the first inverting stage with current bias as stage bias, it is searched for an inverting stage with voltage bias as stage bias \eqref{eq:ainvv}. This type of inverting stage is only part of an op-amp if the first stage fulfills the criteria of a symmetrical op-amp (Line \ref{line:criteriaSymmetricalOpAmp}). It is sufficient to search for a transistor stack $ts_i$ connected with its gate $ts_i.g_1$ to one of the voltage biases in the first stage load. $ts_i$ must be connected with its drain $ts_i.d$ to a voltage bias drain $vb_j.d$. The voltage bias $vb_j$ must be  connected with its gates to the current bias of the already recognized inverting stage $a_{inv,1}$.

As inverting stages might be connected to other inverting stages, as, e.g, a third stage, it is repeatedly iterated over the set of analog inverters until no new stage is found. Thus, also multi-stage op-amps are supported by this method. Multi-stage op-amps usually comprise multiple inverting stage. Some of them may be connected in frequency compensation in feedback loops \cite{MultiStage1, MultiStage2}. However, as they have the characteristic structure of an inverter and as the gate of one of the transistors in the transconductance  is connected to the output of the previous stage, they are unambiguously identifiable.
 
 After finding all amplification stages of an op-amp, the circuit bias, compensation and load capacitors are recognized using their definitions.

\section{Experimental results}\label{sec:experimentalResults}

\begin{figure}[tb] \centering
	\subfloat[HL 2 (Alg. \ref{algo:HierarchlevelTwo})]{
		\label{fig:opAmpCascodeFirstStageHL2}
		\includegraphics[width=0.45\linewidth]{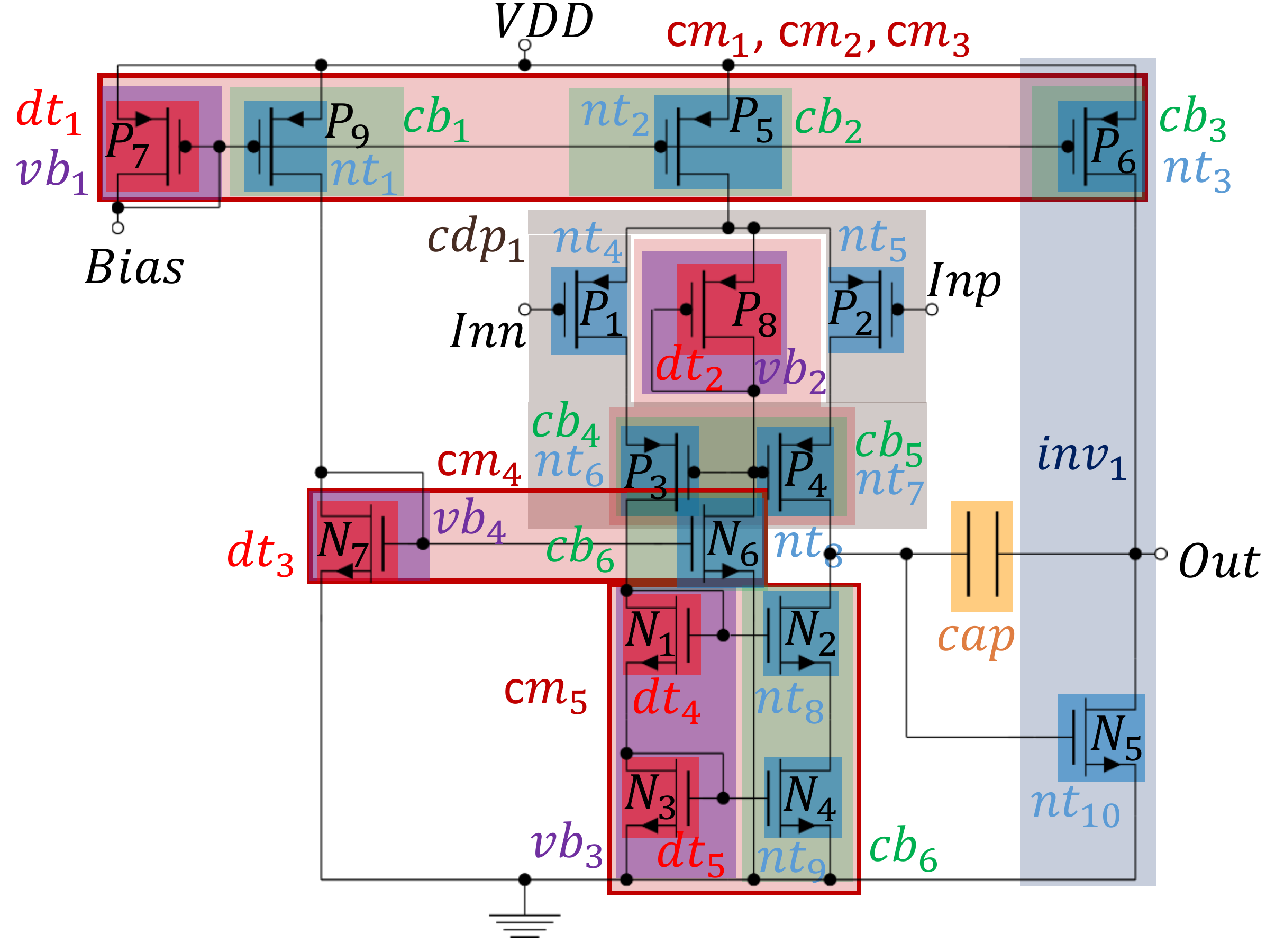}
	}%
	\qquad
	\subfloat[HL 3 - HL 4 (Alg. \ref{algo:HierarchlevelThrreToFive})]{
		\label{fig:opAmpCascodeFirstStageHL34}
		\includegraphics[width=0.45\linewidth]{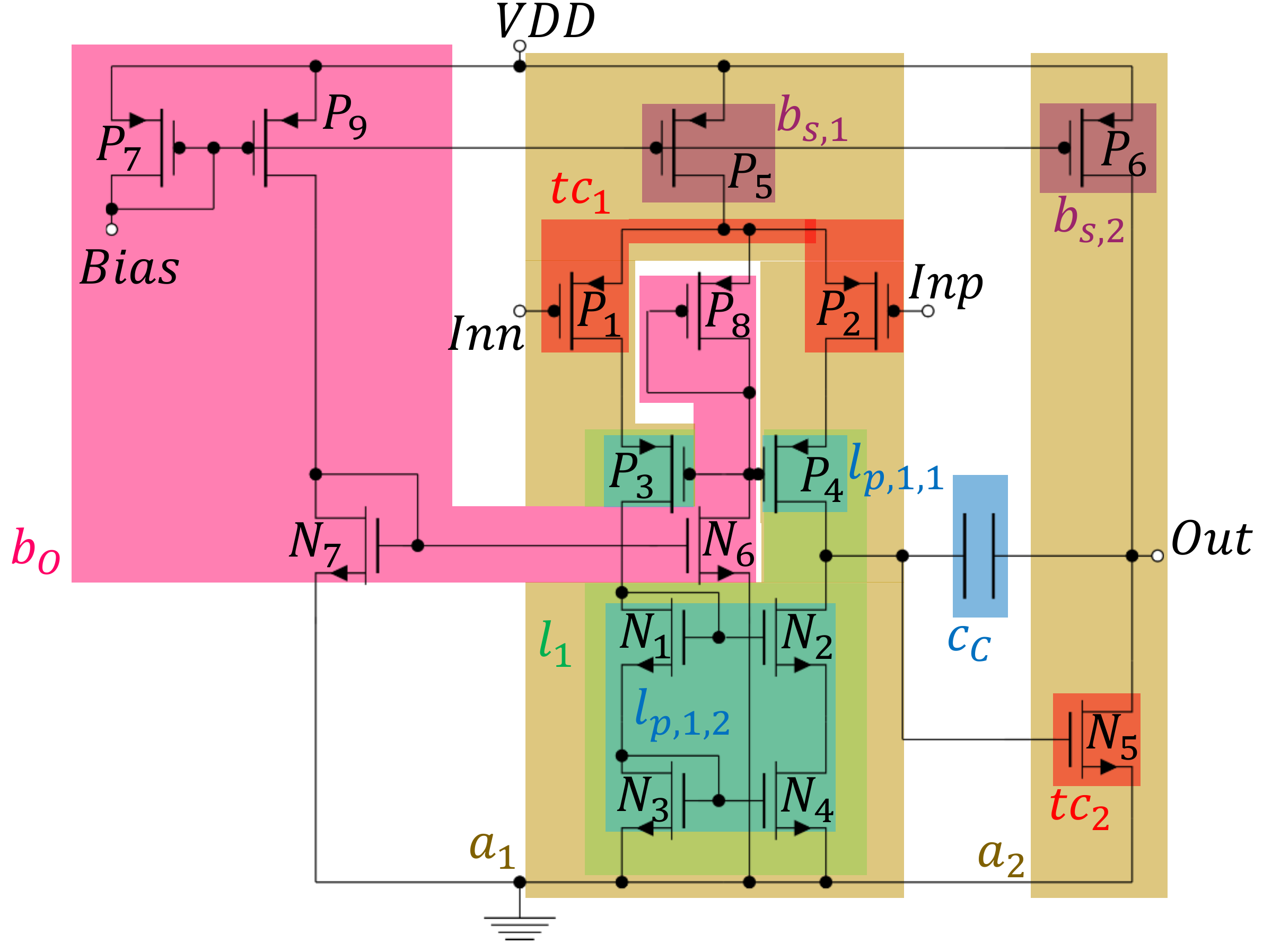}
	}
	\caption{Blocks recognized during functional block analysis in a telescopic two-stage  op-amp}
	\label{fig:opAmpCascodeFirstStage}
\end{figure}

\begin{figure}[t] \centering
	\subfloat[Functional blocks on HL 2]{
		\label{fig:foldedCascodeOpAmpCMFBHL2}
		\includegraphics[width=0.45\linewidth]{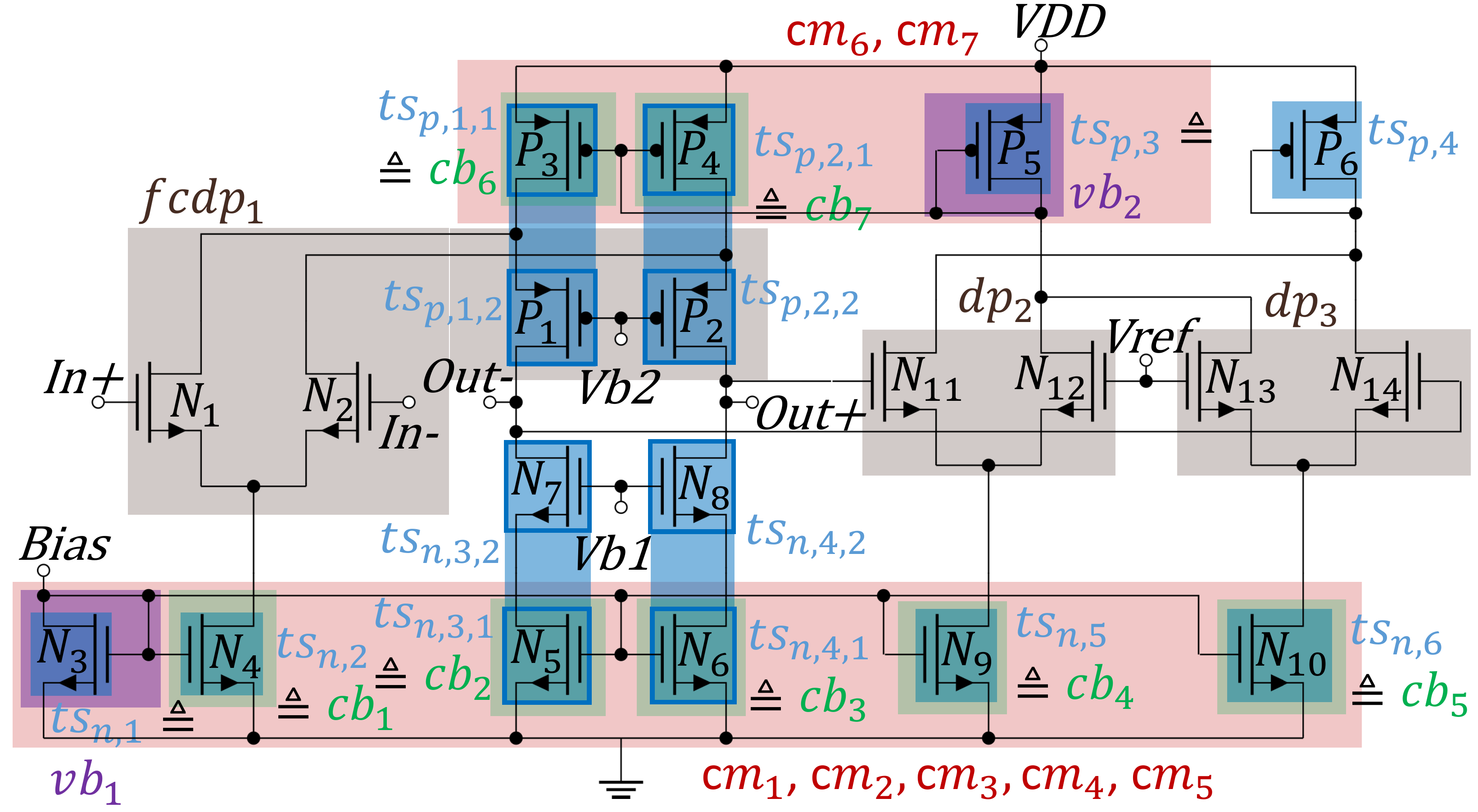}
	}%
	\qquad
	\subfloat[Functional blocks on HL 3 - HL 4]{
		\label{fig:foldedCascodeOpAmpCMFBHL34}
		\includegraphics[width=0.45\linewidth]{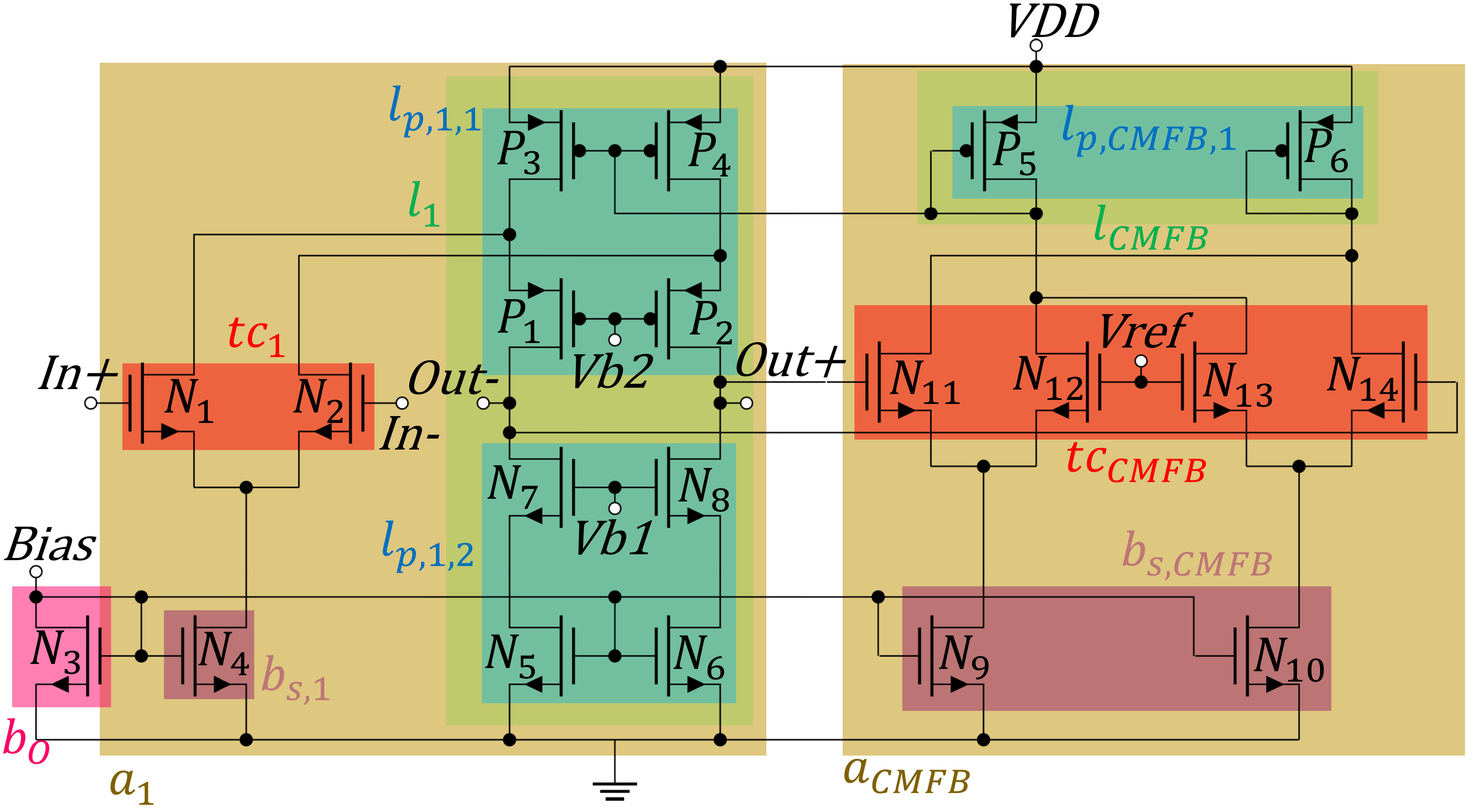}
	}
	\caption{Folded-cascode op-amp with CMFB}
	\label{fig:foldedCascodeOpAmpCMFB}
\end{figure}

\begin{figure}[t] \centering
	\subfloat[Functional blocks on HL 1 - HL 2]{
		\label{fig:threeStageOpAmpHL12}
		\includegraphics[width=0.45\linewidth]{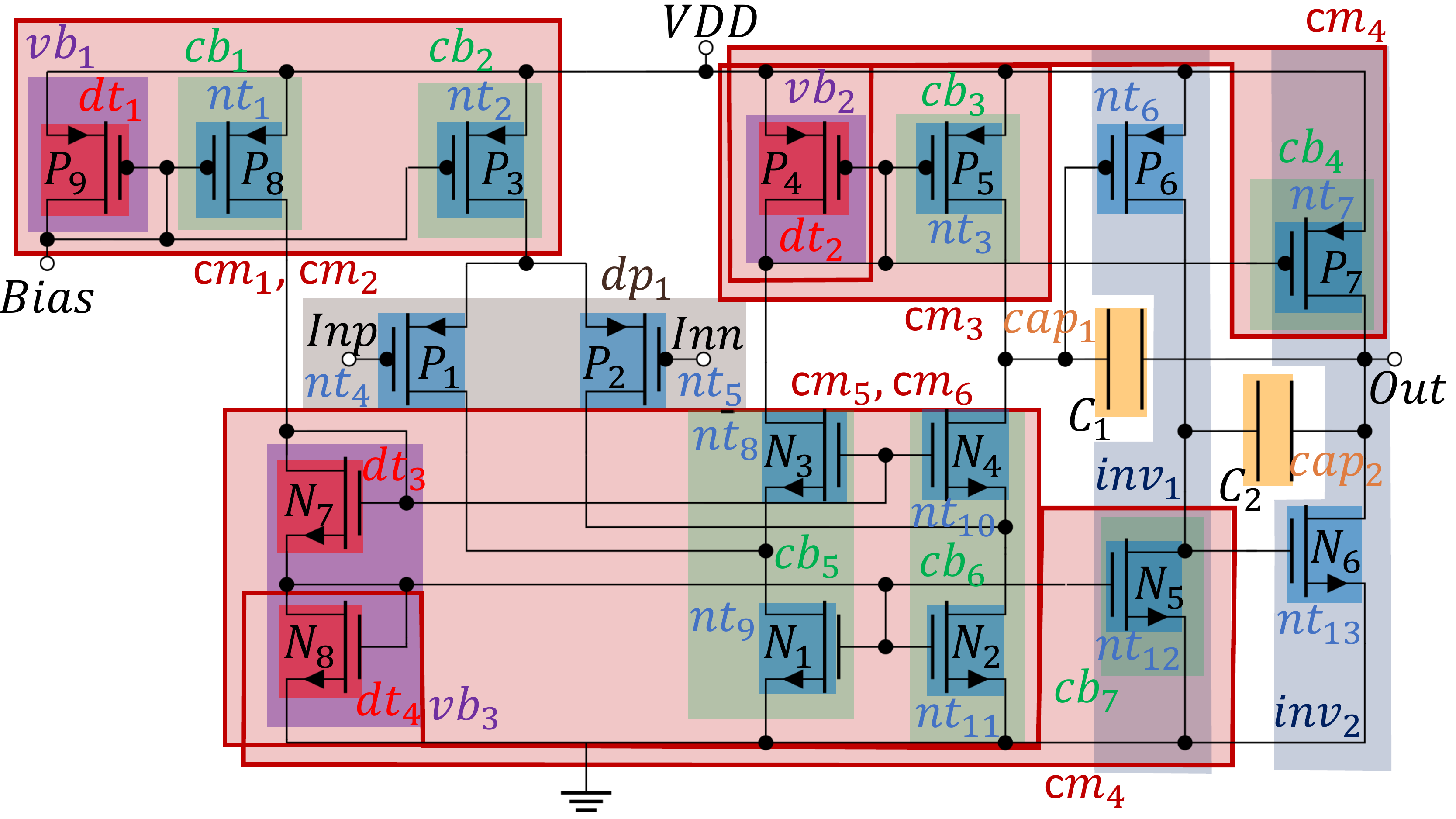}
	}%
	\qquad
	\subfloat[Functional blocks on HL 3 - HL 4]{
		\label{fig:threeStageOpAmpHL34}
		\includegraphics[width=0.45\linewidth]{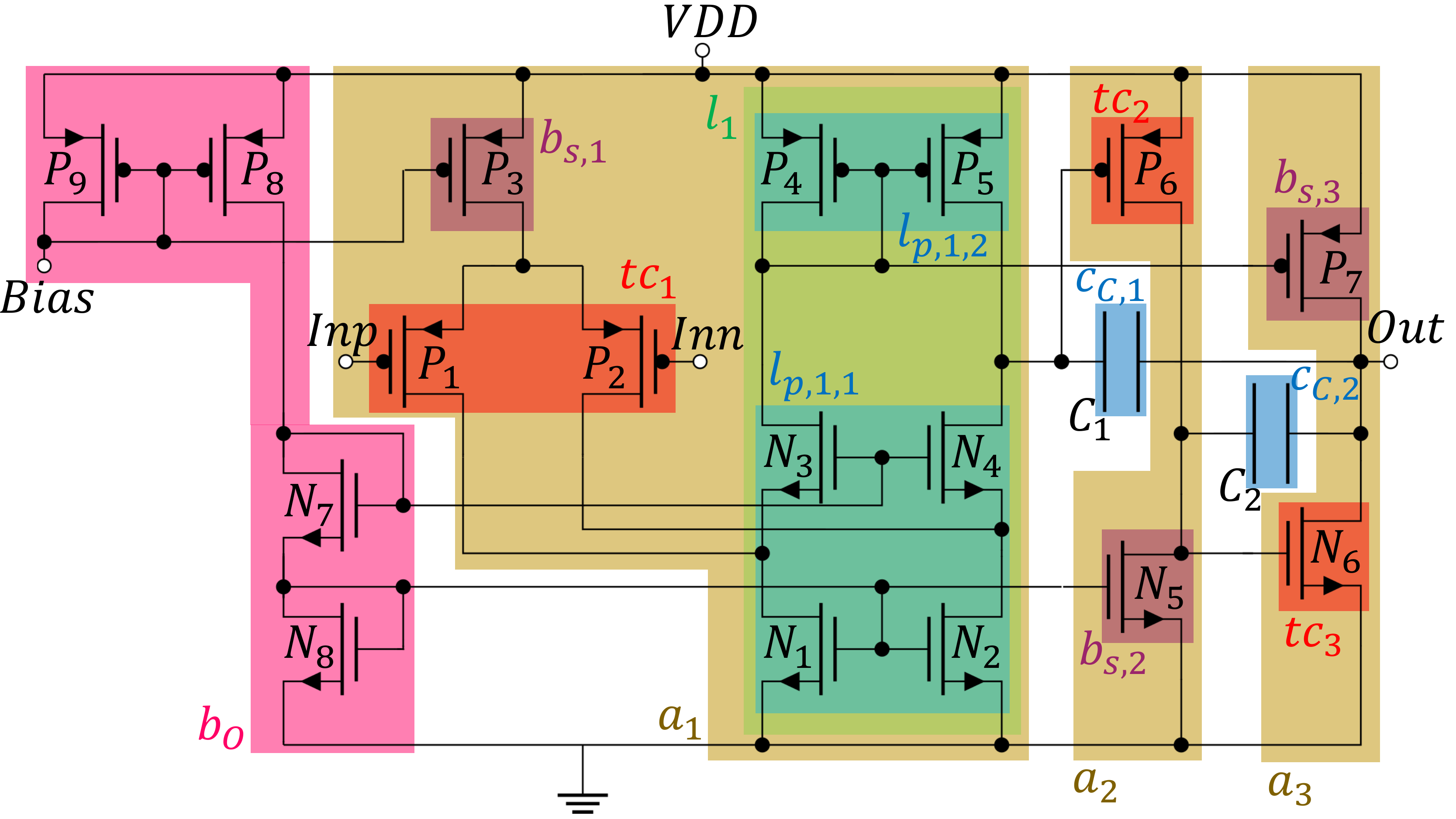}
	}
	\caption{Three-stage op-amp}
	\label{fig:threeStageOpAmp}
\end{figure}

In this section, we present experimental results of the functional block analysis. We illustrate the behavior of the algorithms presented in Sec. \ref{sec:functionalBlockAnalysis} on the example of a telescopic two-stage op-amp (Fig. \ref{fig:opAmpCascodeFirstStage}). Furthermore, we discuss the results of the functional block analysis on three different circuits: a symmetrical op-amp (Fig. \ref{fig:symmetricalOpAmpWithSecondStage}), a folded-cascode op-amp with CMFB  (Fig.~\ref{fig:foldedCascodeOpAmpCMFB}) and a three-stage op-amp (Fig.~\ref{fig:threeStageOpAmp}).
Overall more than 4000 different op-amp topologies have been successfully analyzed.

The computational cost of the algorithm is very low. The runtime for every topology is in the area of milliseconds. The time needed to add a new functional block to the algorithm depends on the uniqueness of its structures. It is very low and in the area of a few hours, if many of the already implemented functional blocks are reused. More advanced functional blocks need a day integration time. Frequency compensation methods as described in \cite{Capacitor1, Capacitor2} can be integrated in this method in less than a work day.

{\em Telescopic op-amp (HL  2):}
Fig. \ref{fig:opAmpCascodeFirstStageHL2} shows all functional blocks recognized on level 2 with Alg. \ref{algo:HierarchlevelTwo}. 
Alg. \ref{algo:HierarchlevelTwo} first searches for all transistor stacks in the circuit sorted according to their substrate type. All transistor stacks are marked in dark blue. Transistor stacks including transistors of the differential pair ($P_1,P_2$) are not shown as they are not valid (Sec. \ref{sec:feasibleAndUnfeasibleAssignments}). Every transistor in the circuit is a transistor stack by itself. 
That means for the three transistor stacks consisting of two transistors, every transistor in these stacks is also a transistor stack by itself. 
The algorithm checks which of the transistor stacks are voltage and current biases. 
Note that, for $P_5, P_8$, the transistor stacks of the individual transistors are  classified as a current bias and voltage bias, respectively. $P_5$ has a gate connection to the voltage bias $vb_1(P_7)$ and therefore fulfills the definition of a current bias. $P_8$ has a gate-drain connection to itself and therefore is a voltage bias connected with its gate to the gates of the current biases $cb_4(P_3)$ and $cb_5(P_5)$.
The other two transistor stacks $N_1,N_3$ and $N_2,N_4$, form a current mirror. The transistor stacks of the individual transistors therefore  are irrelevant.
Note that $vb_2$  does not form  current mirrors with $cb_4$ and $cb_5$ as their sources are not connected to the same net.
All voltage and current biases in this circuit are primary. All voltage biases have their needed gate-drain connection \eqref{eq:vb} by themselves, such that no secondary voltage biases are needed to establish a drain-gate connection of a voltage bias gate. 
Alg. \ref{algo:HierarchlevelTwo} ends with the recognition of the cascode differential pair $cdp_1$ \eqref{eq:cdp} and the analog inverter $inv_1$ \eqref{eq:inv}.

{\em Telescopic op-amp (HL 3 - HL 4):}
Fig. \ref{fig:opAmpCascodeFirstStageHL34} shows all functional blocks recognized on levels 3 and 4 with Alg. \ref{algo:HierarchlevelThrreToFive}.
The algorithm first recognizes  the differential pair as simple non inverting first stage transconductance $tc_1$.  
For the load recognition, the drain nets of $P_3,P_4$ are used.  $P_3,P_4$ form the gate connected couple part of the cascode differential pair $cdp_1$. Two load parts are found:  The gate connected couple forms the load part with p-doping. 
The load part with n-doping consists of the current mirror $cm_5= \{ts_{n,1}, ts_{n,2}\}$.
As stage bias, the current bias $cb_2$ is found as it is connected with its drain to the source of $tc_1$. With it, all parts of the non-inverting first stage $a_1$ are recognized.
In the next step, Alg. \ref{algo:HierarchlevelThrreToFive} checks if the inverter recognized on level 2 is a second stage.
This is true, as $P_6$ was recognized as current bias $cb_3$ and $ts_{n,3}(N_5)$ is connected with its gate to one output of the first stage.
Because the first stage does not fulfill the criteria of a symmetrical op-amp, it is not searched for an inverting stage with voltage bias as stage bias.
The voltage and current biases not part of the two amplification stages are recognized as circuit bias $b_O$.
The recognition ends by identifying the capacitor in the circuit as compensation capacitor.   

{\em Symmetrical op-amp with an additional inverting stage:}
Fig.~\ref{fig:symmetricalOpAmpWithSecondStage} shows the results of the functional block analysis for a symmetrical op-amp with an additional inverting stage $a_3$.
During the recognition of the inverting stages, first $inv_2$ is recognized as inverting second stage $a_{2,1}$. Afterwards $a_{2,2}$ is recognized, $a_{1}$ fulfills the condition of a first stage of a symmetrical op-amp and with $a_{2,1}$, the needed second stage with current bias as stage bias is given \eqref{eq:ainvv}. The transistor stack $\{nt_4(P_3), nt_6(P_5)\}$ fulfills the definition of a non-inverting transconductance $tc_{2,2}$ \eqref{eq:ginv}. $vb_2$ is the stage bias with voltage output. The gate-gate connection to the stage bias of  $a_{2,1}$ is given.
Note, that for a correct identification of $a_3$ the exact definition of an inverting stage must be used \eqref{eq:ainv}. As $P_7$ has a gate connection to one output of the first stage, a drain of the differential pair, it is considered as possible inverting transconductance of another second stage. As for this transconductance, no stage bias exist because $N_6$ is not of type bias, it is not recognized as transconductance. Instead, after $a_{2,1}$ is recognized, $N_6$ is identified as inverting transconductance, because it has a connection to an output of $tc_{2,2}$. $P_7$ is  its stage bias.

{\em Folded-cascode op-amp with CMFB:}
Fig. \ref{fig:foldedCascodeOpAmpCMFB} shows the results of the functional block analysis in a folded-cascode op-amp with CMFB.
The four load transistors $P_1, P_2, N_7, N_8$  are externally  biased by the pins $Vb1$ and $Vb2$ respectively. Hence, these four transistors are not recognized as current biases (Fig.~\ref{fig:foldedCascodeOpAmpCMFBHL2}) as they are not connected with their gates to a voltage bias. However with Alg. \ref{algo:loadRecognition}, they are still recognized as part of the load (Fig. \ref{fig:foldedCascodeOpAmpCMFBHL34}). 
$P_6$ is also neither identified as current nor voltage bias as it does not have any  gate-gate connections to a another transistor with same doping. 
However, it fulfills the functional definition of a load as it has a drain-drain connection to $tc_{CMFB}$.
Therefore, it is recognized with Alg. \ref{algo:loadRecognition} correctly even if it does not fulfill the definition of a load entirely \eqref{eq:l}.

{\em Three-stage op-amp:}
Fig. \ref{fig:threeStageOpAmp} shows the results of the analysis for a three-stage op-amp. Two analog inverters $inv_1, inv_2$ are recognized on HL 2 (Fig. \ref{fig:threeStageOpAmpHL12}). One of the transistors $N_5$ ($inv_1$), $P_7$ ($inv_2$) is identified to be part of a current mirror. The other transistor $P_6$ ($inv_1$), $N_6$ ($inv_2$) is not part of any additional functional block on HL 2.  After recognizing the first  and  second stage with Alg. \ref{algo:HierarchlevelThrreToFive}, the third stage is recognized (Fig. \ref{fig:threeStageOpAmpHL34}). With the gate of $N_6$ connected to the drain of $N_5$, the input of third stage is connected to the output of the second stage \eqref{eq:ainv}. $P_7$ is a identified current bias. Hence, $inv_2$ fulfills all criteria of a third stage. Alg. \ref{algo:HierarchlevelThrreToFive} ends by identifying the compensation capacitors $c_{c,1}, c_{c,2}$. Both capacitors a connected between the output of two stages, $c_{c,1}$ between the output of first and third stage, $c_{c,2}$ between second and third stage. Hence, both capacitors fulfill \eqref{eq:cC}.

\section{Application of the Hierarchical Functional Block Decomposition Method} \label{sec:application}
Two applications of the functional block decomposition method are presented: A sizing method  (Sec. \ref{sec:appSizing}, details in \cite{ABNG20c}) and a structural synthesis method  (Sec. \ref{sec:appSynthesis}, details in \cite{ABNG20d}). 

\subsection{Application to Circuit Sizing} \label{sec:appSizing}
The common manual design process  is based on analytical equations as described in \cite{Allen, LakerSansen, Sansen2007, AnalogIntegratedCircuitDesign, GrayMeyer}.
For each functional block described in Secs. \ref{sec:DeviceLevelFunctionalBlock} - \ref{sec:FunctionalBlockLevel4}, a behavior model can be derived based on these standard equations resulting in a hierarchical performance equation library (HPEL) \cite{ABNG20c}.
For the functional blocks on HL 3, the equations part of the library describe, e.g., the input  and output  conductance ($gin_{fb_k}$, $gout_{fb_k}$) of a functional block $fb_k$. 
Following equation for instance holds for the output conductance $gout_{fb_k}$ of a functional block consisting of one or two transistor stacks $ts_{k}$:
\begin{equation}\label{eq:outputConductanceTS}
	gout_i =\begin{cases}
		gd_{t_{k,out}},~~~~~~~~~~~~~~  \{t_{k,out} \} = ts_{k} \subseteq fb_k \\
		\frac{gd_{t_{k,out}}gd_{t_{k,supply}}}{gm_{t_{k,out}}}, ~~~ \{t_{k,out}, t_{k,supply} \} = ts_k \subseteq fb_k
	\end{cases}
\end{equation}
$gd_{t_{k,j}}, gm_{t_{k,j}}$ are the output und input conductance of a transistor in $fb_k$ with $j$ giving the position of the transistor in the transistor stack.
There are exceptions to this equation which are discussed in \cite{ABNG20c}.

An equation for the output resistance $R_{out,n}$ of an amplification stage functional block $a_n$ on HL 4 is developed based on \eqref{eq:outputConductanceTS} and the $m$ functional blocks of the stage connected to the output net of the stage. These functional blocks can be stage biases, load parts and transconductances.
\begin{equation}\label{eq:outputRestistance}
	R_{out,n} =	\frac{1}{\sum_{j=1}^{m} gout_{fb_j}}
\end{equation}
This equation is part of the equation describing the open-loop gain of a stage $A_{D,n}$ which is on the same hierarchy level in the HPEL: 
\begin{equation}
	 A_{D, n} = gin_n \cdot R_{out,n}; ~~~ A_{D,0} = \prod_{n=1}^l A_{D,n}
\end{equation} 
with $gin_n$ the input conductance of the stage.
The open-loop gain equation of the overall op-amp $A_{D,0}$   is part of HL 5 of the HPEL.
$l$ is the number of all stages in the op-amp, 

\begin{table}[tbp]\scriptsize\centering
	\caption{Results of the sizing method for the telescopic op-amp {\protect Fig. \ref{fig:opAmpCascodeFirstStage}} }\label{tab:outputSizingTool}
	\subfloat[Performance values; M: sizing method; S: simulation]{
		\label{tab:performanceValues}
		\setlength{\tabcolsep}{0.1cm}
		\begin{tabu}{|>{\raggedright\arraybackslash}m{3.7cm}||>{\centering\arraybackslash}m{0.6cm}|>{\centering\arraybackslash}m{0.5cm}|>{\centering\arraybackslash}m{0.5cm}|}
			\hline
			{Constraints} &{ Spec.} & M & S   \\\hline \hline
			Gate-area (10$^3$ $\mu$m$^2$) & $\leq$15  & 5.8& - \\
			Quiescent power (mW) & $\leq$10 & 5.8 &  6.1 \\\hline		
			Max. common-mode input voltage (V) & $\geq$3 &  3.3  & 4.4\\
			Min. common-mode input voltage (V) & $\leq$2 &0 & 0.1\\
			Max. output voltage (V) & $\geq$4   &4.5 & 4.5  \\
			Min. output voltage (V) & $\leq$1 &  0.3  & 0.2 \\
			CMRR (dB) & $\geq$90 & 130  & 146 \\
			Unity-gain bandwidth (MHz) &$\geq$7  & 10 &{7} \\\hline
			Open-loop gain (dB) & $\geq$80 & 120& 93 \\
			Slew rate ($\frac{\text{V}}{\mu \text{s}}$)& $\geq$15 & 28  & 22 \\
			Phase Margin ($^\circ$) & $\geq$60 & 60 & 67   \\\hline
		\end{tabu}
	}%
	\qquad
	\subfloat[Device sizes]{
		\label{tab:DimensionsTelescopicOpAmp}
		\setlength{\tabcolsep}{0.1cm}
		\begin{tabu}{|>{\raggedright\arraybackslash}m{1.6cm}||>{\raggedright\arraybackslash}m{2cm}|}
			\hline
			Variable & Value ($\mu$m/pF)\\\hline \hline
			$W_{P_{1,2}};$$L_{P_{1,2}}$ &	172;9 \\
			$W_{P_{3,4}};L_{P_{3,4}}$&	27;4\\
			$W_{P_5}; L_{P_5}$ &	247;3\\
			$W_{P_6}; L_{P_6}$ &	515;3\\
			$W_{P_7}; L_{P_7}$ &	7;3\\
			$W_{P_8}; L_{P_8}	$& 7;4\\
			$W_{P_9}; L_{P_9}$ &	43;3\\
			$W_{N_{1,2}};$$L_{N_{1,2}}$ &	90;1\\
			$W_{N_{3,4}};$$L_{N_{3,4}}$&	90;1\\
			$W_{N_5}; L_{N_5}$ &	130;1\\
			$W_{N_6}; L_{N_6}$&	269;1\\
			$W_{N_7}; L_{N_7}$&	166;9\\
			$C_c$ &	6.4\\
			\hline
		\end{tabu}
	}%
	\qquad
\end{table}

Analogously, performance equations for all functional blocks that have been presented in Sec. \ref{sec:DeviceLevelFunctionalBlock} - Sec. \ref{sec:FunctionalBlockLevel4} have been established \cite{ABNG20c}. After a functional block analysis of a given op-amp topology according to the methods presented in the paper, the corresponding performance equations of the functional blocks are automatically composed forming a behavioral model of the op-amp performance. This model forms a Mixed-Integer Non-linear Programming problem (MINLP) for sizing the circuit, which is solved, e.g., with a constraint programming approach \cite{ABNG20b}.

Table \ref{tab:outputSizingTool} shows the results of the sizing method for the telescopic op-amp (Fig. \ref{fig:opAmpCascodeFirstStage}) using the specifications given in column 2 of Table Ia. The values calculated with the analytical equations within the sizing method are compared to simulation results using a BSIM3v3 models. The deviations due to the simpler transistor model are less then 30\% and within designer expectations.

\subsection{Application to Structural Synthesis}\label{sec:appSynthesis}
The functional block library (Sec. \ref{sec:DeviceLevelFunctionalBlock} - Sec. \ref{sec:FunctionalBlockLevel4}) can be applied to structural synthesis by functional block composition. Alg. \ref{algo:SynthesisFunctionalBlock} gives an algorithm to create the structural implementations $S_{new}$ of a functional block $fb_{new}$ based on the implementation $S_1, .., S_i$ of its subblocks and a set of characteristic connections $R_c$.
The two-transistor implementations of a current bias  (Fig. \ref{fig:currentBiases}), e.g., are created based on two structural implementation sets $S_1, S_2$ containing  normal and diode transistors of the same doping $\Phi$ (${S}_1: NT_\Phi, DT_{\Phi};$ ${S}_2: NT_\Phi;$). Every combination of  $s_1 \in S_1, s_2 \in S_2$ has a drain-source connection (${R}_c: s_1.d \leftrightarrow s_2.s$). Thus, all  structural implementations of a two-transistor current bias are created.
More complex functional blocks need additional rules set as explained in \cite{ABNG20d}.

\begin{algorithm} [tp]{ \small
		\caption{{\small Synthesis of a functional block}} \label{algo:SynthesisFunctionalBlock}
		\begin{algorithmic}[1]
			\REQUIRE Set of subblock implementations $S_1, .., S_i$; Connection rules $R_c$
			\STATE ${S}_{new}$ := $\{ ~~\}$ //{\small The set of structural implementations of ${fb}_{new}$ is empty}
			\FORALL{$s_1 \in {S}_1$}
			\STATE ....
			\FORALL {$s_i \in {S}_i$}
			\STATE $s_{new}$ := createNewImplementation($s_1, ..., s_i, {R}_c$)
			\STATE ${S}_{new}$ := ${S}_{new} \cup \{s_{new}\}$
			\ENDFOR
			\STATE ...
			\ENDFOR
			\RETURN ${S}_{new}$
	\end{algorithmic}}
\end{algorithm}

With Alg. \ref{algo:SynthesisFunctionalBlock}, the structural implementations of all functional blocks in Secs.~\ref{sec:DeviceLevelFunctionalBlock} - \ref{sec:FunctionalBlockLevel4} are created allowing a creation of up to 3000 topologies.
The corresponding composition graph is presented in \cite{ABNG20d}.
For given specifications, the topologies are sized and evaluated using an enhanced version of the sizing method described in Sec. \ref{sec:appSizing}. The respective algorithm is presented in \cite{ABNG20d}.
Table \ref{tab:structureResults} shows the output of the synthesis tool for the specifications in Table Ia. As the specifications are not highly demanding and no upper bounds are specified, many topologies are able to fulfill the specifications, e.g. the telescopic op-amp in Fig. \ref{fig:opAmpCascodeFirstStage}. Defining more strict specifications reduces the number of created topologies significantly.

\begin{table}[tbp]
	\scriptsize\centering	\setlength{\tabcolsep}{0.1cm}	
	\caption{Amplification stage composition of the resulting topologies for the specification in Table Ia}\label{tab:structureResults}
	\begin{tabular}{|>{\raggedright\arraybackslash}m{2.0cm}||>{\centering\arraybackslash}m{0.4cm}|>{\centering\arraybackslash}m{0.3cm}||>{\centering\arraybackslash}m{0.8cm}|>{\centering\arraybackslash}m{0.7cm}||>{\centering\arraybackslash}m{0.6cm}|>{\centering\arraybackslash}m{0.6cm}||>{\centering\arraybackslash}m{1.7cm}||c|}
		\hline	
		First stage type &\multicolumn{2}{>{\centering\arraybackslash}m{1cm}||}{ simple} & \multicolumn{2}{>{\centering\arraybackslash}m{1.8cm}||}{folded-cas\-code} & \multicolumn{2}{>{\centering\arraybackslash}m{1.3cm}||}{tele\-scopic} & sym\-met\-ri\-cal & \multirow{2}{0.8cm}{total } \\ \cline{1-8}
		\# stages & 1 & 2 & 1 & 2 & 1 & 2 & - &\\ \hline \hline
		 \# topologies & 0 &	111& 0&181& 20&77&58&	447 \\ \hline
	\end{tabular}
\end{table}

\section{Conclusion and Outlook}\label{sec:conclusion}
This paper presented a new method to represent op-amps by their functional block. The functional blocks are ordered hierarchically. For each functional block, a systematic functional and structural description is given allowing an automatic recognition.
Two different areas of applications are presented: 
Together with a performance equation library, the functional block analysis allows a topology-independent sizing of op-amps.
A functional block composition method allows structural synthesis of op-amps. Thousands of different op-amp topologies are created based on hierarchically generated functional blocks by small rule sets.

Future work remains in including more feedback compensation techniques in the method. Currently, a compensation capacitor is supported in the recognition algorithm. For multi-stage op-amps, additional compensation circuits as, e.g., \cite{Capacitor1, Capacitor2} are needed. As they have a fixed structure, the compensation circuits can be added to the method. 
Future research will also be on including the functional block composition method in layout synthesis.

\bibliography{custom_added_2}

\end{document}